\documentclass[a4paper,11pt]{quantumarticle}
\pdfoutput=1

\usepackage{amsmath,amssymb}
\usepackage{algpseudocode}

\usepackage{newfloat}
\DeclareFloatingEnvironment[name=Algorithm,placement=tbp]{algorithm}
\algrenewcommand\algorithmiccomment[1]{\unskip\hfill\hskip 1em$\triangleright$~#1}
\newenvironment{algobox}[1][]{\begin{algorithm}\caption{#1}\small}{\end{algorithm}}
\newenvironment{algobox*}[1][]{\begin{algorithm*}[t]\caption{#1}\footnotesize}{\end{algorithm*}}


\usepackage{microtype}
\usepackage{mathrsfs}
\usepackage{xcolor}
\usepackage{graphicx}
\graphicspath{{figures/}}
\usepackage{bm}
\usepackage[hidelinks]{hyperref}
\makeatletter
\AtBeginDocument{\renewcommand*{\theHALG@line}{\thealgorithm.\arabic{ALG@line}}}
\makeatother
\usepackage{cleveref}
\usepackage{placeins}

\crefname{section}{Sec.}{Secs.}
\Crefname{section}{Section}{Sections}
\crefname{figure}{Fig.}{Figs.}
\Crefname{figure}{Figure}{Figures}
\crefname{table}{Table}{Tables}
\Crefname{table}{Table}{Tables}
\crefname{equation}{Eq.}{Eqs.}
\Crefname{equation}{Equation}{Equations}
\crefname{algorithm}{Algorithm}{Algorithms}
\Crefname{algorithm}{Algorithm}{Algorithms}

\usepackage{amsthm}
\usepackage{braket}
\usepackage{multirow}
\usepackage{booktabs}
\usepackage{makecell}

\theoremstyle{definition}
\newtheorem{definition}{Definition}

\title{A Scalable, Optimized Scheduler for the Active Volume Architecture}

\author{Sam Heavey}
 \email{sheavey@psiquantum.com}
 \affiliation{PsiQuantum, 700 Hansen Way, Palo Alto, CA 94304, USA}
\author{Athena Caesura}
 \email{acaesura@psiquantum.com}
 \affiliation{PsiQuantum, 700 Hansen Way, Palo Alto, CA 94304, USA}

\begin{document}
\maketitle

\begin{abstract}
    \noindent We improve the accuracy of Active Volume (AV) resource estimates by explicitly scheduling when AV blocks execute and when their reactions complete. We present a cycle-accurate scheduler that assigns each logical qubit a role in every logical cycle. The approach allows us to capture execution effects such as gate packing inefficiencies, parallelization overhead, and memory requirements. In addition, the scheduler explicitly tracks reaction latency and stale state lifetimes, rather than assuming that stale states can be absorbed into a fixed overhead. For a $4\times4$ Fermi--Hubbard test circuit, we find a transition to reaction-dominant regime at approximately $10^3$ logical qubits. Beyond $10^3$ logical qubits, stale state accumulation restricts the effective workspace and the benefit of adding more qubits is substantially decreased. Below this regime the scheduler predicts a $1.64\times$ runtime speedup relative to the previous Analytic Resource Estimators (ARE), primarily due to a $1.80\times$ reduction in average memory occupancy. We also show that parallelization and reaction latency overhead grows with available workspace, inverting the trend of the ARE model used in \cite{Caesura25}. These results provide a path toward reaction-aware analytic resource estimates and a full compilation pipeline for AV architectures. We also develop novel techniques for estimating Y state and routing demand and apply them to our test circuit. The Y state estimates result in small efficiency boost and the routing demand estimates show that routing should not bottleneck early fault-tolerant AV devices.

\end{abstract}


\section{Introduction}

\noindent As the era of early fault-tolerant quantum computing draws closer, the quantum computing community has produced increasingly refined resource estimates. The first resource estimates for quantum computation counted the gates and qubits required to implement a given algorithm~\cite{beckman1996efficient, beauregard2002circuit}. As the theory of quantum error correction developed and quantum architectures became better defined, resource estimates began to report time-to-solution and physical-qubit counts as the main cost metrics~\cite{litinski2019game, gidney2021factor, litinski2023compute, Caesura25}. However, these resource estimates suffer from a lack of detail, merely counting the most relevant gates and their costs. In recent literature, we have seen resource estimates that attempt to actively schedule each gate rather than just counting them~\cite{saadatmand2024fault, leblond2023tiscc}. 

In this paper, we apply this new class of resource estimates to the Active Volume (AV) architecture, which is particularly well suited to photonic quantum computing~\cite{Litinski22Active}. We introduce a compiler, the \emph{block scheduler}, that assigns qubits to roles in each logical cycle and greedily minimizes the number of logical cycles. The scheduler explicitly accounts for qubit overheads and reaction latency required in parallel execution at each logical cycle. To keep the approach broadly applicable across AV architectures, the block scheduler deliberately abstracts away spatial implementation details such as qubit routing and nonuniform decoder throughput. We assess the block scheduler by performing quantum resorce estimates (QREs) on a representative early fault-tolerant application on a photonic quantum computer and compare it to the analytic methods used in~\cite{Caesura25}.

The block scheduler enables larger circuits to be executed on a given hardware configuration and allows us to test the memory-allocation assumptions of~\cite{Litinski22Active}, particularly the fraction reserved for tasks beyond AV execution and quantum information storage. Our resource estimates also account for often-overlooked overheads associated with qubit bridging, $\ket{Y}$ state management, and stale states. From these results, we derive new analytical models for bridge-qubit and stale-state accumulation.

Stale-state accumulation can lead to the reaction-limited regime, in which delayed reactions slow logical execution and increase hardware requirements~\cite{Khalid2025DecoderLatency, gidney2019flexible}. Previous resource estimates assumed that reactions would be sufficiently fast to avoid this regime during early fault-tolerant quantum computation~\cite{Caesura25, Litinski22Active}. We reproduce the exponential accumulation of stale states predicted in~\cite{Khalid2025DecoderLatency} and demonstrate its impact on runtime. We find that the onset of reaction-limited behavior depends on the algorithm, computer size, and hardware parameters. For a qubitized Fermi–Hubbard circuit executed on a photonic architecture, it can occur at approximately 900 qubits. Conversely, our analytical model identifies reasonable early fault-tolerant hardware parameters under which reaction depth does not affect runtime.

Finally, we use the block scheduler to quantify routing demand. We show that on our test circuit a computer of modest size only must rearrange 20\% of the qubits at each logical cycle, well within the assumptions made in~\cite{Litinski22Active}. We also define two routing paradigms, intra- and inter-cycle routing and show that routing doesn't necessarily scale with computer size. For computations where the majority of qubits are used as workspace, the routing overhead remains small.

We begin in \cref{sec:active_volume} by introducing the Active Volume architecture and refining the terminology for the qubit roles used throughout the paper. In \cref{sec:compiler_description}, we present the block scheduler, including the directed acyclic graph (DAG) construction and the greedy scheduling algorithm. In \cref{sec:resource_estimation}, we explain how resource estimates are obtained both analytically and with the block scheduler. Finally, in \cref{sec:results}, we use these tools to test and refine assumptions from~\cite{Litinski22Active}, quantify bridge- and stale state overheads, identify the regime in which reaction times become limiting, and benchmark the scheduler’s runtime performance, optimality, and scalability.

\section{Active Volume Architecture}
\label{sec:active_volume}

\noindent Quantum gates are applied to qubits (space) and take time to be implemented. For this reason, every quantum gate consists of some minimal spacetime footprint, which we measure in units of spacetime blocks (blocks). The number of blocks and the precise cost of a block depends on the fault-tolerant (FT) protocol used to encode the qubits and implement gates. In this paper, we shall consider the surface code and implement gates through lattice surgery on a photonic quantum computer, but note that AV works for any architecture that has sufficiently fast qubit routing~\cite{Litinski22Active}. Gates implemented via lattice surgery on distance $d$ surface code patches require one logical cycle, defined as $d$ syndrome measurements. Therefore, $d$ parametrizes the height of a spacetime block (the time axis) while, for square surface code patches, the two spatial dimensions are $d\times d$~\cite{Litinski22Active, fowler2012surface}.

\begin{figure*}[p]
    \centering
        \includegraphics[width=.9\textwidth]{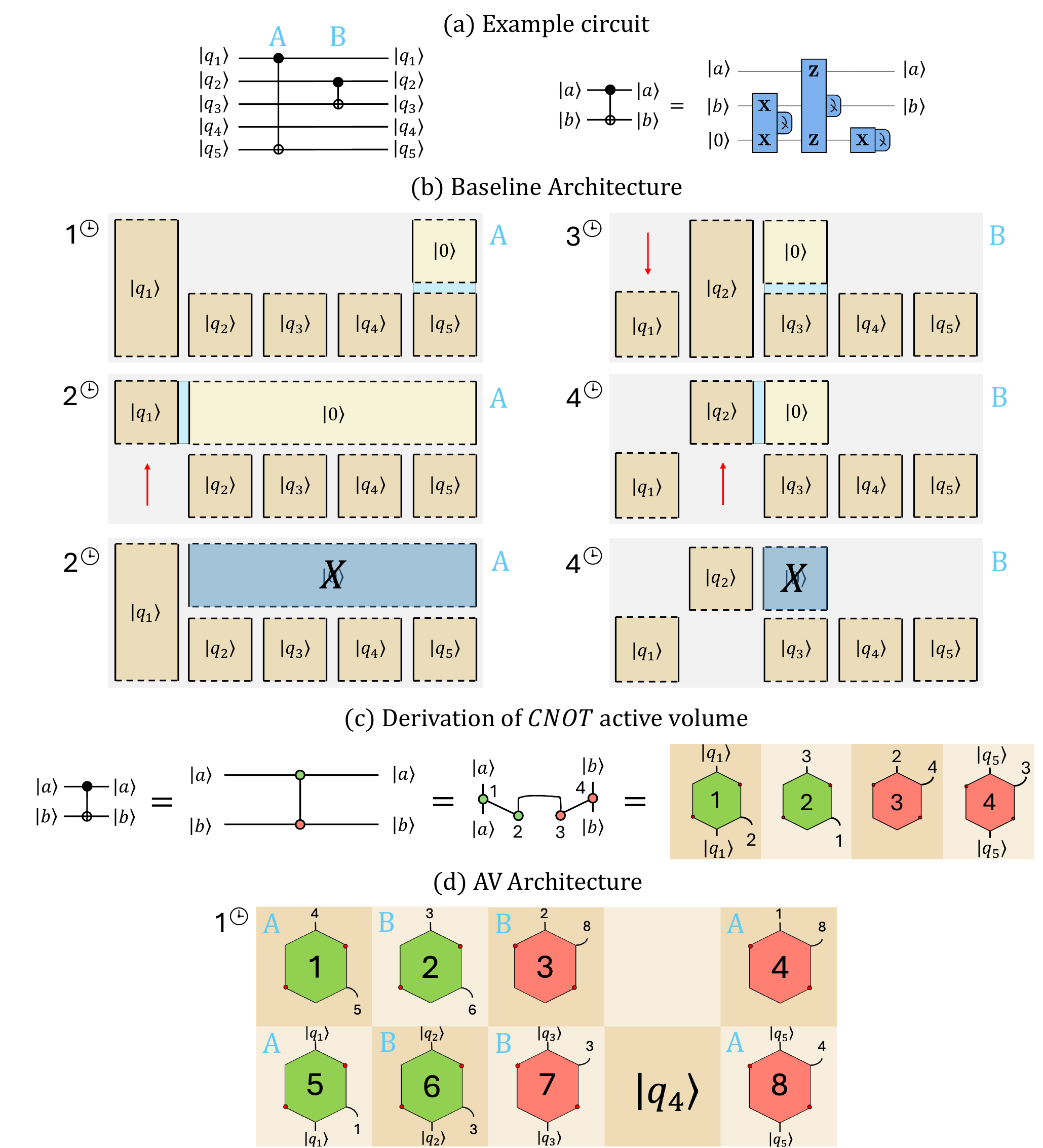}
    \caption{a) A circuit that consists of two $CNOT$ gates, ${CNOT}_A$ and ${CNOT}_B$, acting on disjoint sets of qubits (left). A $CNOT$ gate written as a series of Pauli-Product-Measurements (PPMs) (right). Note Pauli corrections are omitted, since they can be tracked classically with no physical gate cost. b) An architecture where qubits exist as a 2D array of surface code patches that have nearest-neighbor-only connections, described in detail in \cite{litinski2019game}. The dashed (solid) boundaries represent the logical X (Z) operator for each patch. The blue patches with an overlaid $X$ indicate destructive $X$-basis measurements on all physical qubits. The diagram flows in order of increasing time steps (which have units of logical cycles) such that the first column shows the execution of ${CNOT}_A$ between $\ket{q_1}$ and $\ket{q_5}$ and the second column shows ${CNOT}_B$ between $\ket{q_2}$ and $\ket{q_3}$. The computation employs 10 qubits over 4 logical cycles which equates to a total spacetime volume of 40 blocks. c) The AV derivations of a $CNOT$ gate using surface code qubits. We start by writing both gates as ZX diagrams (middle-left) and then convert these into orientated ZX diagrams (OZX) (middle-right). We then formally write the OZX diagram in logical block notation, counting 4 blocks. Both OZX diagrams and logical block notation have a one-to-one correspondence with spacetime blocks which represent the cost of lattice surgery between surface codes, for more details on OZX diagrams and logical block notation see~\cite{Litinski22Active}. d) We execute the circuit from a) on a 10 qubit computer in the AV architecture, due to long-range connections both gates can be completed in one logical cycle, with a qubit to spare. Logical blocks 1,4,5,8 implement ${CNOT}_A$ and 2,3,6,7 implement ${CNOT}_B$. The total spacetime cost of this computation is 10 qubits multiplied by 1 logical cycle which equals 10 blocks.
}
    \label{fig:baseline_vs_av}
\end{figure*}

In an AV architecture, we allow for long-range routing between qubits \cite{Litinski22Active}. Photonic quantum computers are particularly well suited to facilitate such connections as it has been shown that an arbitrary level of connectivity between qubits can be achieved by adding more switch options and fiber optic delays \cite{bombin2021interleaving}. However, adding many long-range connections increases the hardware complexity and error rates, and is likely unnecessary \cite{Litinski22Active}. Therefore, a photonic AV architecture should minimize long-range connectivity between qubit modules.

Long-range connections allow us to easily execute gates in parallel and reduce qubit idling. To elucidate this fact, we execute a toy circuit in both the AV architecture and an architecture consisting of a static 2D array of qubits with nearest-neighbor-only connections (baseline), see \cref{fig:baseline_vs_av}. In the baseline architecture, we must stretch our patches such that the correct boundaries are brought together for lattice surgery. This can take up a significant amount of space on the computer and block the execution of other gates. The first two time steps of \cref{fig:baseline_vs_av}(b) are spent executing a $CNOT$ between $\ket{q_1}$ and $\ket{q_5}$. Because these qubits are far apart, we require a large ancillary patch to connect them. This consumes a lot of space and prevents us from executing the $CNOT$ between $\ket{q_2}$ and $\ket{q_3}$ in parallel. In total, it takes $4$ time steps (logical cycles) to execute the gates on the $10$-qubit baseline computer. We therefore incur a spacetime cost of $4 \times 10 = 40$ blocks. However, \cref{fig:baseline_vs_av}(c) shows that a $CNOT$ costs 4 blocks; the gates minimum spacetime cost—the active volume (AV)—is only 8 blocks. Therefore, 80\% of the allocated computational volume is idle. In~\cref{fig:baseline_vs_av}(d) we execute the circuit again but in the AV architecture. Here, we relax connectivity constraints and are able to complete both gates in just $1$ logical cycle. In fact, we could even reduce the computer size to $9$ qubits as one patch is completely unused. Doing so means we have just one qubit, $\ket{q_4}$ idling during the gate. This gives us an idle volume of $11.1\%$ with no waste, which is a $68.9\%$ reduction in wasted volume relative to the baseline implementation. It is precisely because we no longer need to stretch ancillary patches that we can execute a gate with its minimum spacetime footprint. Note that you could rearrange qubits in~\cref{fig:baseline_vs_av}(b) such that the ancillary patch is shortened but moving patches takes time and so still adds to the spacetime cost of the computation. In the AV architecture, qubits can be easily rearranged so they become adjacent, avoiding this cost altogether.

\subsection{Terminology}
\label{sec:terminology}
\begin{figure}
    \includegraphics[width=\linewidth]{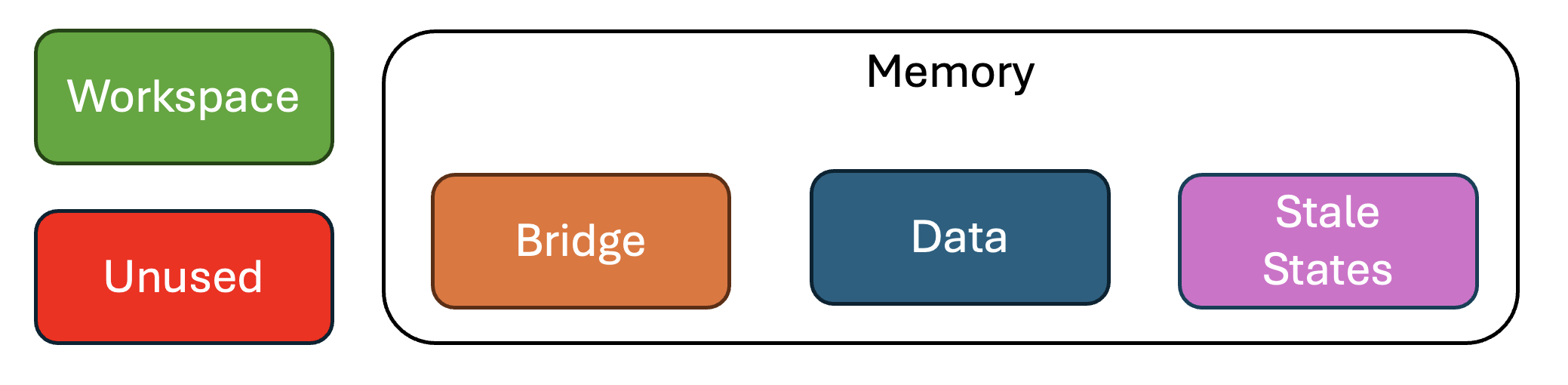}
    \caption{A diagram showing the nested definitions of different qubit types. The goal of our block scheduler will generally be to increase the workspace size as much as possible, while decreasing the number of unused and bridge qubits.}
    \label{fig:qubit_classifications}
\end{figure}

\noindent In this section, we make explicit the qubit classifications outlined in~\cite{Litinski22Active}. A summary of the nested definitions is given in~\cref{fig:qubit_classifications}. Qubits in the AV architecture are assigned to either workspace, memory, or unused.

\begin{definition}[Memory qubit]
    \emph{Memory qubits} are idling to maintain coherence, so they are free to be rearranged during the execution of a logical cycle.
\end{definition}
\begin{definition}[Workspace qubit]
    A \emph{workspace qubit} executes 1 Active Volume block in 1 logical cycle. Workspace qubits cannot be rearranged during the logical cycle because they are executing lattice surgery gates.
\end{definition}
\begin{definition}[Unused qubits]
    \emph{Unused qubits} refer to idle qubits that would ideally be workspace qubits, but due to constraints of the scheduler, must be left idle. Unused qubits can be rearranged during a logical cycle. They were treated as insignificant in the AV paper~\cite{Litinski22Active}, so this is the only new term introduced here.
\end{definition}

Next, we define three subtypes of memory qubits:

\begin{definition}[Data qubits]
A \emph{data qubit} is either an input or output qubit in a circuit diagram. These include magic states.
\end{definition}

\begin{definition}[Stale states]
A \emph{stale state} is any qubit that sits idly while a decoder determines the basis in which it should be destructively measured (reactive measurement).
\end{definition}

\par \textbf{Stale states.} Gates implemented via measurements (e.g., in fusion-based architectures~\cite{bartolucci2023fusion}) require corrections when an undesired measurement outcome is recorded. We care about two classes of corrections: Pauli and non-Pauli corrections. Pauli corrections do not have to be physically implemented and can instead be used to update our interpretation of measurement outcomes. Non-Pauli corrections, which are generated by non-Clifford gates, must be physically implemented either by directly applying the correction or by using a stale state. Both of these methods require processing measurement outcomes to check if a correction is needed. The first method stalls the computation, idling many qubits, so we choose to use stale states and reactive measurements. Stale states trade space for time by creating a new qubit, the stale state, which can effectively defer the correction choice by encoding it into a later measurement basis choice. Crucially, the computation can continue while the decoder determines a measurement basis for the stale state, which is now the only qubit idling. See~\cref{sec:magic_states} for more details on stale states and the reaction time.

\begin{definition}[Bridge qubits]
A \emph{bridge qubit} is the qubit from a Bell state that idles during a logical cycle while its sister qubit is used as the input for an AV block.
\end{definition}

\par
\begin{figure}
    \includegraphics[width=\linewidth]{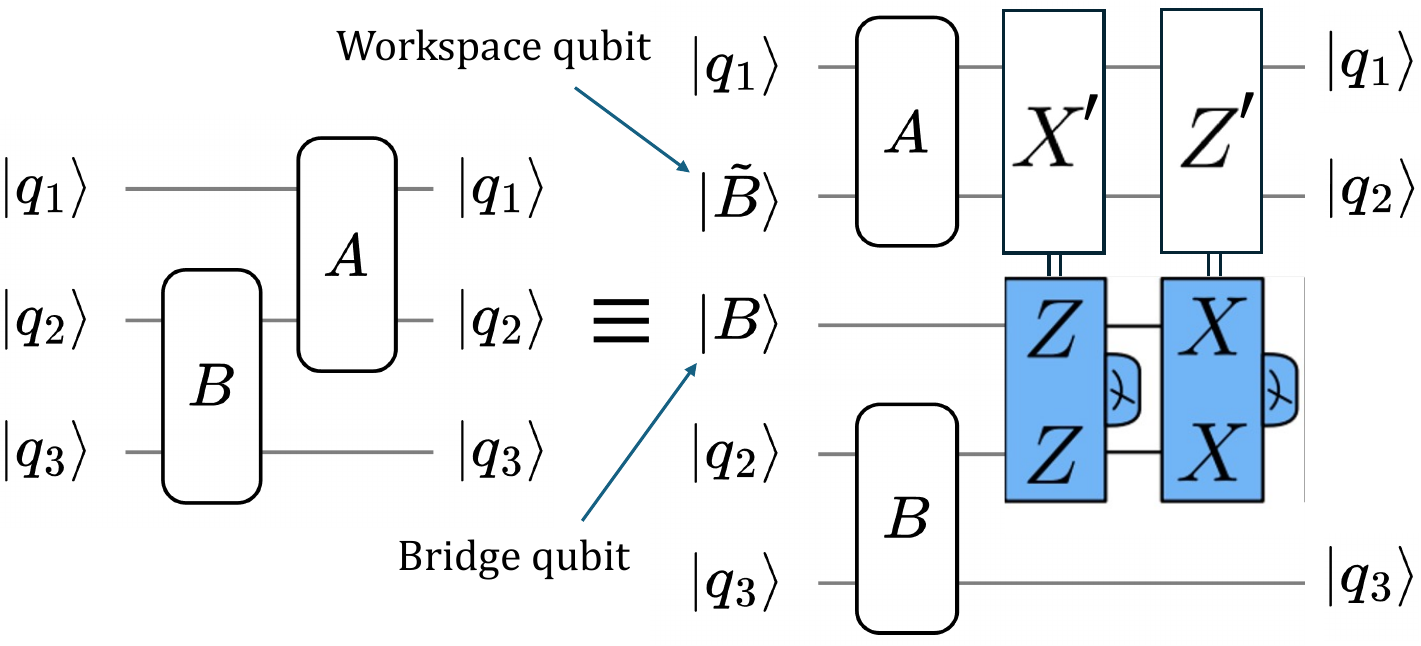}
    \caption{Gates $A$ and $B$ share qubit $\ket{q_2}$. To execute $A$ and $B$ in parallel (right) we initialize the Bell state $\ket{B\tilde{B}}$ (e.g., $\frac{1}{\sqrt{2}}\left[\ket{00}+\ket{11}\right]$). $\ket{\tilde{B}}$ is used in place of $\ket{q_2}$ as an input into $A$, while $\ket{B}$ sits idle. After $A$ and $B$ are executed, we perform a Bell measurement on $\ket{q_2}$ and $\ket{B}$. A Pauli $X$ or $Z$ correction is tracked depending on the $ZZ$ and $XX$ measurement outcomes, respectively. These corrections are transformed by $A$ into $X'=A^{\dagger}XA$ and $Z'=A^{\dagger}ZA$. Note, $\ket{q_2}$ may also represent a qubit register, if that is the case then this process is applied to each qubit in the register (i.e., one Bell state for each qubit).}
    \label{fig:bridge_qubits}
\end{figure}

 \textbf{Bridge qubits.} Consider attempting to execute two gates in parallel that share one or more data qubits (i.e., within the same logical cycle), each shared data qubit can only be used as an input to one of the gates. This problem is solved by initializing a Bell state (two maximally entangled qubits). We place one qubit in memory, which we refer to as a \textit{bridge qubit}, and place the other in workspace. The workspace qubit is used as the input to one of the gates, in place of the shared data qubit. At the end of the cycle, we perform a Bell measurement between the bridge qubit and the original (now transformed) data qubit. This teleports the original data qubits logical state onto the substituted path, so the computation reflects both gates. The resulting logical state appears at the output location of the gate whose input was substituted, see~\cref{fig:bridge_qubits}. Note that bridging therefore increases the required memory for a logical cycle and so reduces the potential size of the workspace.

\subsection{Summary of Prior AV Resource Estimation Methods}
\label{subsec:prior_ARE_methods}
 
\noindent There have been several attempts to calculate the runtime of various quantum algorithms in the AV architecture~\cite{Litinski22Active,litinski2023compute,Caesura25}. We will refer to them collectively as Analytic Resource Estimates (ARE). AREs have historically followed the prescription:
\begin{enumerate}
    \item Set the number of logical qubits by picking a code distance. 
    \item Sum the AVs for each gate in the computation.
    \item Set the number of memory qubits required by the computation using the qubit high-water, that is, the maximum number of data qubits that exist at the same time during the computation.
    \item Increase the number of memory qubits by some factor (e.g., 20\%) for the additional memory required to store all magic states, stale states, and bridge qubits necessary throughout the computation.
    \item Allocate the remaining qubits in the computer to the workspace.
    \item Divide the total AV by the number of workspace modules to get the number of logical cycles. If this is too large for the picked code distance, return to step 1.
    \item Multiply the duration of a logical cycle by the number of logical cycles to get the total runtime. 
\end{enumerate}

\par This runtime estimate is simple and quick to compute, but suffers from three main drawbacks. Firstly, the 20\% memory allocation is only supported by a small study done in~\cite{Litinski22Active} on an adder circuit. In this paper, we aim to refine this estimate by properly accounting for qubit roles other than data or workspace in the model. Bridge qubits in particular are unlikely to follow this pattern since more bridge qubits will be used as the workspace size increases and more gates are done in parallel rather than serially. Second, the prevalence of unused qubits has not been quantified precisely. Although \cite{Litinski22Active} includes a hand-worked small-scale example, a comprehensive accounting across an entire computation on a realistic circuit has been missing prior to this work. Lastly, the ARE methods do not model the effect of reaction times. They assume the existence of a sufficiently fast decoder for stale state removal such that stale states do not accumulate and are accounted for by the $20\%$ factor.

\section{Compiler Description}
\label{sec:compiler_description}

\noindent The goal of our block scheduler is to improve the accuracy of previous resource estimates by properly accounting for stale states, bridge qubits, data qubits, discrete gate fitting, and reaction latency. 
The scheduler we present here assumes the following:
\begin{enumerate}
  \item \textbf{Global logical cycles.} All active volume blocks being executed concurrently must start and finish their execution at the same time. 


  \item \textbf{Qubit routing.} Qubits in memory can be rearranged by using quick-swap operations so there are no delays between logical cycles. These quick-swaps are enabled by long-range fiber optic connections.

  \item \textbf{Fusion graphs.} Lattice surgery may only occur between qubits that are less than $r$ modules away. We assume $r$ is large enough for the execution of all gates. Note, \cite{Litinski22Active} estimates $r=12$ to be sufficient.
  


  \item \textbf{Deterministic magic state distillation.} The distillation protocols we use for $\ket{T}$ and $\ket{CCZ}$ magic state distillation are probabilistic. Here we assume distillation is always successful, but use AV costs from~\cite{Litinski22Active} that factor in failure rates.
\end{enumerate}

\subsection{DAG Creation}
\label{sec:dag}
\noindent The block scheduler, as its name implies, is designed to address the scheduling of gates. Scheduling problems are naturally represented by a directed acyclic graph (DAG), where vertices denote gates and directed edges encode whether one gate occurs before another. Concretely, we assign each gate in the computation a vertex and insert an edge $u \to v$ whenever gate $u$ must precede gate $v$. In this work, we use a DAG to address two compilation tasks.

First, following the prescription given in~\cref{sec:terminology}, we use stale states to implement corrections for non-Clifford gates. These stale states must be measured in appropriate bases, and the choice of basis for a given measurement may depend on prior measurement outcomes. Consequently, stale state measurements satisfy a partial ordering that can be captured using a DAG~\cite{ruh2025quantum}. By layering this DAG (i.e., counting the minimal number of edges that must be traversed before an gate becomes executable) we can quickly compute the~\emph{reaction depth}. Reaction depth is the number of sequential basis-update layers in the full circuit, called~\emph{reaction layers}. Counting layers rather than individual updates gives a far more realistic estimate, since independent updates may be carried out in parallel. The time needed to determine the next basis from the current one is the~\emph{reaction time}. In this paper, we primarily use this framework to measure the number of reaction layers required within a single logical cycle, rather than the total reaction depth of the full computation, since our focus is on whether reaction constraints increase the logical cycle length.

Second, we use the DAG representation to quantify bridging overhead. By tracking the qubit masks associated with each gate, we can compute pairwise qubit overlap between gates that are candidates for concurrent execution and thereby estimate the number of bridge qubits required to realize that concurrency. This bridging cost is central to the block scheduler's objective: it seeks, when possible, to minimize the number of bridge qubits, since they consume resources that could otherwise be allocated to executing AV (workspace).

\begin{figure}[!tbp]
\centering
\includegraphics[width=.99\linewidth]{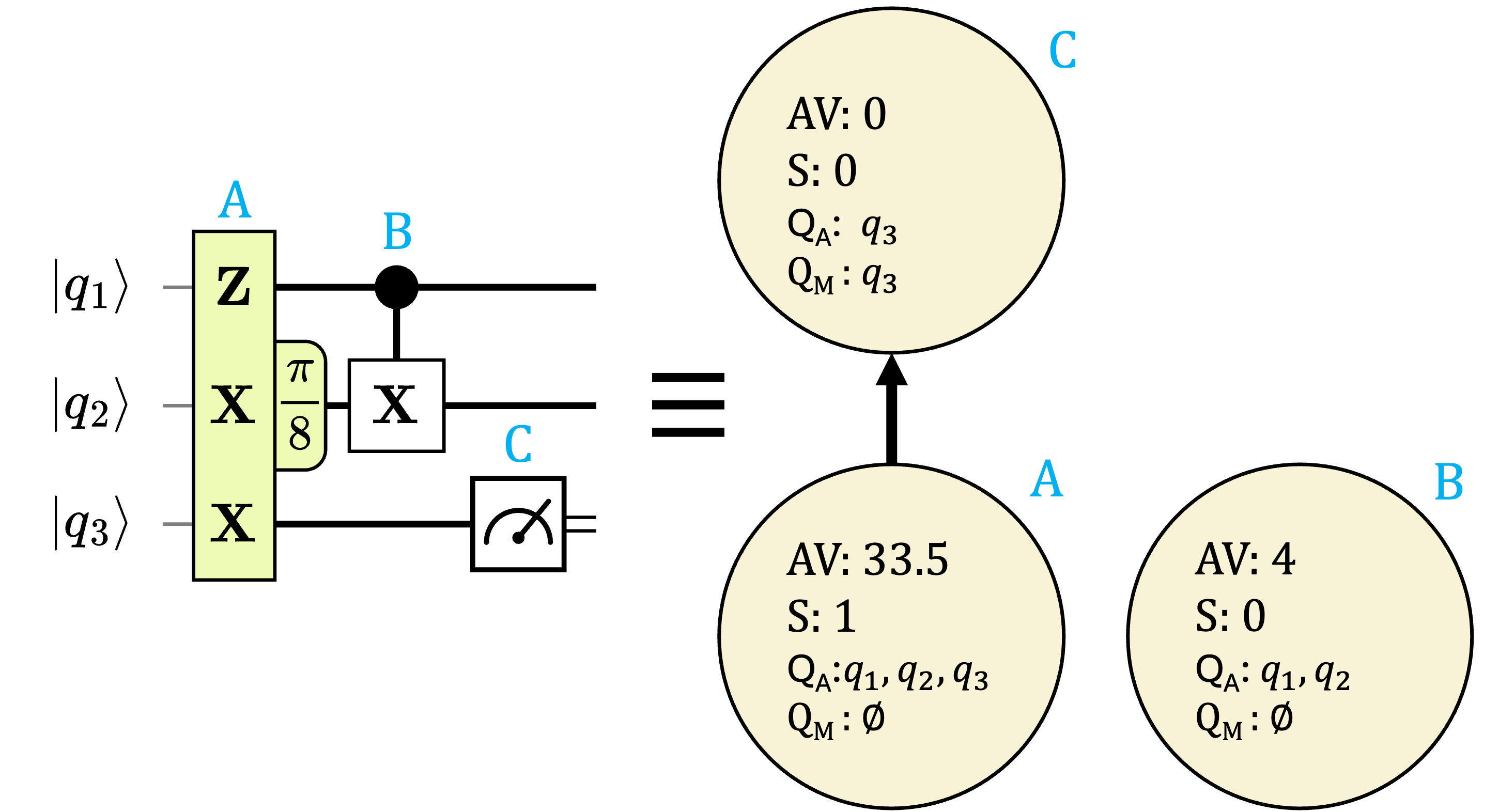}
\caption{Conversion of a simple quantum circuit into a DAG. Each vertex stores the qubits acted on, the stale state count, $\ket{Y}$-state consumption, and active volume. The attributes of vertex $A$ were obtained from the $\frac{\pi}{8}$ PPR decomposition in~\cref{fig:magic_state_injection}.}
\label{fig:circuit_dag_conversion_example}
\end{figure}

To support both tasks, we construct a DAG in the following way. Let $G = (V, E)$ be a DAG, where $V$ is the set of vertices representing gates and $E$ is the set of directed edges between the gates. For example, if two non-commuting gates, $A$ and $B$, act on at least one of the same qubits and $A$ comes before $B$, then a DAG, $G$, will show $A$ with an arrow pointing to $B$ ($A,B \in V$ and $(A,B) \in E$). If $A$ and $B$ act on disjoint sets of qubits or commute, then they are not connected by an arrow, see~\cref{fig:circuit_dag_conversion_example}.

Pseudocode for our DAG construction procedure is given in \cref{alg:dag_backward_search}. This presentation is intentionally high-level and omits implementation details, most notably the caching data structures we use to accelerate DAG construction in practice. Even with these optimizations in place, we will see in~\cref{sec:results} that DAG construction remains the primary bottleneck in our overall compilation time.

\begin{algobox}[Construct Operation DAG]
\label{alg:dag_backward_search}
\begin{algorithmic}[1]
\Require Ordered gates $\mathcal{O}=(o_1,\dots,o_m)$
\Ensure Directed acyclic graph $G=(V,E)$

\State Initialize directed graph $G$ with vertices $V=\{1,\dots,m\}$ and $E\gets\emptyset$
\State $Leaves \gets \emptyset$ \Comment{frontier / sink set}

\ForAll{$v \in V$}
  \State $Q \gets \Call{List}{Leaves}$ \Comment{backward-search queue}
  \State $F \gets Leaves$ \Comment{seen set for queue membership}

  \State $i \gets 1$
  \While{$i \le |Q|$}
    \State $u \gets Q[i]$
    \State $i \gets i+1$

    \If{\Call{Commutes}{$o_u,\,o_v$}}
      \ForAll{$p \in \mathrm{Pred}(u)$}
        \If{$p \notin F$}
          \State append $p$ to $Q$
          \State $F \gets F \cup \{p\}$ 
        \EndIf
      \EndFor
    \Else
        \State add edge $(u \rightarrow v)$ to $G$
        \State $Leaves \gets Leaves\setminus\{u\}$
    \EndIf

  \EndWhile



  \State $Leaves \gets Leaves\cup\{v\}$
\EndFor

\State \Return $G$
\end{algorithmic}
\end{algobox}

We define the following properties which will be used for DAG construction and the block scheduler. For each vertex $v \in V$:
\begin{align*}
\mathrm{AV}(v) &:\ \text{Active volume in units of blocks}, \\
\mathrm{S}(v) &:\ \text{Stale state count}, \\
\mathrm{Q_A}(v) &:\ \text{Set of qubits acted on or initialized}, \\
\mathrm{Y}(v) &:\ \text{$\ket{Y}$ consumption probabilities,} \\
\mathrm{\bar{Y}}(v) &:\ \text{Number of catalyst $\ket{\bar{Y}}$ used,} \\
\mathrm{Q_M}(v) &:\ \text{Set of qubits measured}, \\
\mathrm{Pred}(v) &:= \{\,u\in V \mid (u,v)\in E\,\}, \text{ and} \\
\mathrm{Desc}(v) &:= \{\, u \in V \mid \exists \text{ a directed path } v \leadsto u\,\}.
\end{align*}

Consider an example schedule based on the DAG in \cref{fig:circuit_dag_conversion_example}. Suppose gates $A$, $B$, and $C$ are assigned to the same logical cycle. The scheduler must verify that
\begin{align}
\mathrm{AV}(A) + \mathrm{AV}(B) + \mathrm{AV}(C) + b \le w,
\end{align}
where $w$ is the available workspace and $b$ is the bridge qubit overhead, ensuring that the scheduled gates fit. If the constraint is satisfied, the corresponding vertices occupy two DAG layers: $A$ and $B$ in the first layer, followed by $C$ in the second. Consequently, if $C$ produces stale states, they are removed after two reaction times. In the next logical cycle, the workspace must be reduced by the number of stale states produced in the current logical cycle. Here, $\mathrm{S}(A)=1$, so the workspace is reduced by one qubit \footnote{When enough non-commuting stale states are produced within a logical cycle, the reaction times required to resolve them can exceed the logical-cycle duration. These stale states then persist into the next logical cycle, where they can add to the stale-state layers generated in that cycle. A rough calculation for a photonic quantum computer suggests that the threshold is approximately $25$ reaction layers; see \cref{tab:hardware_parameters}.}.
Next, $\mathrm{Q_A}(A)=\{q_1,q_2,q_3\}$, $\mathrm{Q_A}(B)=\{q_1,q_2\}$, and $\mathrm{Q_A}(C)=\emptyset$. Since $q_1$ and $q_2$ are shared by $A$ and $B$, executing this logical cycle requires two bridge qubits, so $b=2$. Moreover, $\mathrm{\bar{Y}}(A)=\mathrm{\bar{Y}}(B)=\mathrm{\bar{Y}}(C)=0$, since none of the gates directly require catalyst $\ket{\bar{Y}}$ states. Furthermore, $\mathrm{Y}(B)=\mathrm{Y}(C)=0$. However, $A$ is a $\frac{\pi}{8}$ Pauli product rotation with a $50\%$ chance of requiring a reactive $Y$ measurement, so $\mathrm{Y}(A)=0.5$ and $A$ may require a $\ket{Y}$ state. Finally, since $\mathrm{Q_M}(C)=\{q_3\}$, $q_3$ ceases to be a data qubit after this logical cycle, freeing it for other tasks.

\subsubsection{Magic states}
\noindent Magic states are created via distillation. Distillations do not have predecessors and so appear in the 0th layer of a DAG. If they are not given special treatment by a scheduler they may be executed long before a magic state is required. As magic states take up space, we may run into a situation where a scheduler prematurely fills the computer with magic states, which then slows or stops a computation.

Luckily, there is an efficient solution, in which schedulers do not have to treat distillation differently from other gates. We pair distillations and gates, where a distillation points to the gate that implements its magic state. We then insert this pair into the DAG at the position where the gate should be, such that predecessors of the gate now point to the distillation and the gate points to its children (no change). Now it is likely the distillation will be scheduled in the same logical cycle or in the logical cycle before the gate. If distillation is scheduled in the logical cycle before the gate then we incur no additional cost due to magic state storage, if it is scheduled in the same logical cycle we require one additional block of bridge qubits per qubit in the magic state, if distillation is scheduled two logical cycles before the gate we must store each qubit in the magic state for one logical cycle and so incur the same block cost as same cycle distillation.

\subsubsection{Y states}
\label{sec:y_states}
\noindent Catalyst states ($\ket{\bar{Y}}$) are commonly used to implement phase gates~\cite{jones2012layered, Litinski22Active}. We assume the computer starts with one $\ket{\bar{Y}}$ state, which serves as the catalyst for producing additional $\ket{Y}$ states. When a gate requires a catalytic $\ket{Y}$ state, we provide $\ket{\bar{Y}}$ directly, or bridge $\ket{\bar{Y}}$ if it is already in use. We distinguish two types of $\ket{Y}$ state consumption: 

1) Deterministic consumption. Here, $\ket{Y}$ states are used directly in gate logic (e.g., an $S$ gate teleportation/injection circuit with a $\ket{Y}$ ancilla). To model this cost, we initialize a $\ket{Y}$ state factory in the workspace and produce $\ket{Y}$ states as needed. All but the first $\ket{Y}$ state produced can be attached to gates in the same logical cycle without additional bridge qubits by flipping Bell state preparations (D port connections for readers familiar with OZX diagrams) to Bell measurements (U port connections) in the $\ket{Y}$ factory (see~\cite{Litinski22Active} Fig.~13(c)).

2) Probabilistic consumption. Here, $\ket{Y}$ states are consumed probabilistically as corrections conditioned on prior measurement outcomes (e.g., a reactive $Y$ measurement after $\ket{T}$ state injection, which requires a $\ket{Y}$ state with 50\% probability). We considered two accounting models: (i) a deterministic expected-value model, which allows fractional $\ket{Y}$ state usage, and (ii) a stochastic simulation model, which consumes only whole $\ket{Y}$ states according to measurement outcomes. We adopt the stochastic model because it better represents real computations. Over long computations, many such events occur, so simulated runtimes fluctuate only slightly around the average. To support this, we store the number of $\ket{Y}$ states and their consumption probabilities as attributes of DAG vertices.

\subsection{The Block Scheduler}
\noindent After converting an algorithm into DAG form, we can place each gate into a logical cycle.  The placement of each gate will determine the number of qubits in each of the 5 roles defined in~\cref{sec:terminology}. We wish to find an assignment of gates that minimizes the number of logical cycles. There are a few factors to consider for an efficient assignment: 
\begin{itemize}
    \item We must respect the memory requirements of each logical cycle and ensure our scheduled gates fit into the remaining unused qubits. 
    \item Respect gate ordering. Let $\mathrm{Desc}(v)$ denote the set of descendants of a vertex $v$. If $v$ is assigned to logical cycle $i$, then for all $u \in \mathrm{Desc}(v)$ we have $\quad i \le \ell (u)$. Where $\ell (u)$ denotes the logical cycle in which the descendant $u$ is executed.
    \item Performing gates in parallel. When gates that share qubits are placed into the same logical cycle, we must initialize a Bell state for each shared qubit. This reduces the amount of unused qubits available for scheduling other gates.
    \item Discrete gate fitting. The AV of a gate must be added to the workspace in its entirety, which means each logical cycle may have some unused qubits when there are no ops small enough to fit into the remaining space. We should try to minimize unused qubits by finding a gate schedule that maximizes workspace each logical cycle.
    \item Non-Clifford gates require magic states which leave behind stale states. These stale states will take up space until they are reactively measured. If we schedule a long chain of DAG gates into one logical cycle, we may create stale states whose reaction times are on the order of a logical cycle. This breaks assumption 2 of the assumptions listed at the beginning of~\cref{sec:compiler_description}.
\end{itemize}

The solution space of valid assignments of gates to logical cycles grows exponentially with algorithm size. Solutions can be found using integer linear programming (ILP), but finding optimal solutions for large algorithms requires significant resources~\footnote{We originally intended to employ an algorithm which used ILP methods, but after implementing the ILP method we found that the greedy method was far more practical. Even hybrid solutions didn't seem to provide enough benefit to justify the decrease in scalability.}. Instead, we developed a greedy algorithm that obeys the restrictions imposed by points 1 and 2 and is guided by the following heuristics:
engage as many different data qubits as possible to minimize bridging,
prioritize vertices with many descendants to spread out reaction depth, and decompose high-AV DAG vertices into multiple low-AV vertices. The last point is motivated by the expectation that, on average, many low-AV vertices generate fewer unused qubits than a small number of high-AV vertices. 

\begin{algobox*}[Greedy scheduling]
\label{alg:greedy-block-scheduler}
\begin{algorithmic}[1]
\Require DAG $G=(V,E)$, $n_l$ and $n_l$ the number of reaction layer left for a stale state and cleared per logical cycle, respectively. $\mathrm{React}$ determines the $\ket{Y}$ production cost expected for non-decoded stale states.
\State $\ell \gets 0$ \Comment{logical cycle index}
\State $D\gets \emptyset$ \Comment{set of existing data qubits}
\State $\Sigma \gets \emptyset$ \Comment{multiset of remaining active stale states}
\State $R \gets \{\, v \in V \mid \mathrm{Pred}(v)=\emptyset \,\}$
\State $\mathcal{R} \gets \mathrm{Sort}(R)$ \Comment{non-increasing by \# descendants}
\While{$\mathcal{R} \neq \emptyset$}
  \State $\mathcal{L_{\ell }} \gets \emptyset$ \Comment{scheduled vertices}
  \State $Q \gets \emptyset$ \Comment{set of data qubits in workspace}
  \State $\Sigma_f \gets \{\, s \mid (s,n_l)\in\Sigma,\ n_l \le 2n_r \,\}$
  \Comment{stale states decoded next logical cycle}
  \State $U \gets \text{total qubits} - |D\setminus Q|$ \Comment{memory requirements}
  \State $U\gets U-|\Sigma|$ \Comment{active stale states}
  \State $U \gets U-\mathrm{React}(\Sigma_f)$ \Comment{$\ket{Y}$ cost for reactive $Y$ demand}
  \State $\kappa \gets 0$ \Comment{allowed qubit overlap count}
  \While{$\kappa \le$ total qubits}
    \State $added \gets \textbf{false}$
    \ForAll{$v \in \mathcal{R}$}
        \State $C_{react} \gets 0$
        \If{$\mathrm{S}(v)>0$}
        \State $\mathrm{depth}(v) \gets 1+ \max\{\, \mathrm{depth}(p) - n_r(\ell-\ell_p) \mid p \in \mathrm{Pred}(v) \,\}_+$
        \If {$\mathrm{depth}(v)\le n_r$} \Comment{Make reactive $\ket{Y}$s now}
        \State $C_{react} \gets \mathrm{React}(\mathrm{S}(v))$
        \EndIf
        \EndIf
        \State $O \gets \, \mathrm{Q_A}(v) \cap Q$ \Comment{set of overlapping qubits}
        \State $W \gets U + \left|\mathrm{Q_A}(v) \setminus O\right|$ 
        \If{$|O| \le \kappa$
          \textbf{and} $\mathrm{AV}(v) + |O| + C_{react} \le W$}
        \State $\mathcal{L_{\ell }} \gets \mathcal{L_{\ell }} \cup \{v\}$ \Comment{schedule $v$}
        \State $U \gets U - \mathrm{AV}(v) - |O|$ \Comment{update $U$}
        \State $\mathcal{R} \gets \mathcal{R} \setminus \{v\}$
        \State Update $\mathcal{R}$: add vertices whose \textit{only} unscheduled predecessor is $v$
        \State $Q \gets \bigl(Q \cup \mathrm{Q_A}(v)\bigr) \setminus \mathrm{Q_M}(v)$
        \State $D \gets \bigl(D \cup \mathrm{Q_A}(v)\bigr) \setminus \mathrm{Q_M}(v)$
        \State $added \gets \textbf{true}$
      \EndIf
    \EndFor
    \If{\textbf{not} $added$}
      \State $\kappa \gets \kappa + 1$ \Comment{Allow for more bridging}
    \EndIf
  \EndWhile
  \State $\Sigma \gets \{\, (s,n_l-n_r) \mid (s,n_l)\in\Sigma,\ n_l>n_r \,\}$
  \State $\Sigma \gets
      \Sigma \uplus
      \{\, {(s_i,\mathrm{depth}(v)) \mid s_i \in \mathrm{S}(v)} \,\}$
      \Comment{add stale states created by $v$}
  \State $\ell \gets \ell + 1$ \Comment{advance to next logical cycle}
\EndWhile
\end{algorithmic}
\end{algobox*}

\begin{figure*}[tp]
    \centering
    \hspace*{-.35cm}%
    \includegraphics[width=1\textwidth]{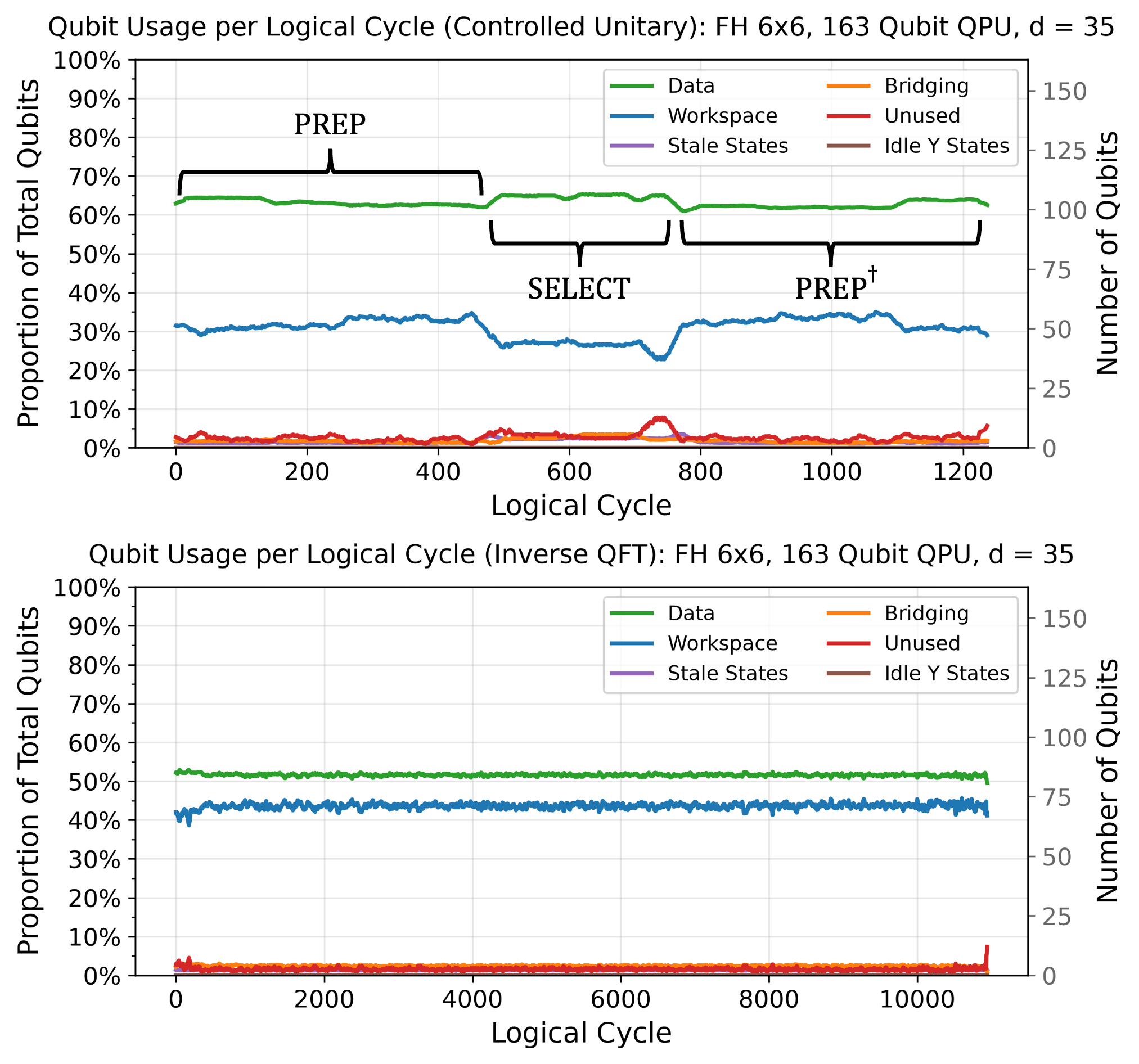}
    \caption{Qubit usage per logical cycle for simulations of a qubitized Fermi--Hubbard model on a $6\times6$ lattice is shown for the two primary subcircuits: the controlled unitary (top) and the QFT$^\dagger$ (bottom). To better reveal usage trends, we apply a rolling average to each qubit category over a 25-cycle window. In both cases, the numbers of stale states and bridge qubits remain well below the 20\% threshold assumed in~\cite{Caesura25} throughout the simulation. We plot idle $\ket{Y}$ states explicitly, even though they are data qubits, to show that our method for handling reactive $Y$ measurements (see \cref{subsec:greedy_algorithm}) does not cause significant buildup of these states in memory. In the controlled unitary, the \texttt{PREP}, \texttt{SELECT}, and \texttt{PREP}$^\dagger$ stages are clearly visible: the number of data qubits increases to accommodate qubits introduced by left-elbow gates~\cite{gidney2018halving} in the \texttt{SELECT} step. One might expect that, since the cost of the QFT is dominated by rotations, the number of bridge qubits would be higher. But we instead find that the number of bridge qubits is lower than in the controlled unitary case. Lastly, spikes in unused qubits identify algorithmic bottlenecks where high-AV gates struggle to fit within the computer, thereby highlighting potential sources of optimization.}
    \label{fig:qubit_usage_per_logical_cycle}
\end{figure*}

\subsubsection{A Greedy Algorithm for Block Scheduling}
\label{subsec:greedy_algorithm}
\noindent We summarize our greedy algorithm using pseudocode, but first we define the $\mathrm{Sort}$ function and discuss reactive $\ket{Y}$ measurements: 
\paragraph{Sorting by descendant count.}
Given $R\subseteq V$, let $\mathrm{Sort}(R)=(r_1,\ldots,r_{|R|})$ be an ordering of the vertices in $R$ such that
\[
|\mathrm{Desc}(r_i)| \ge |\mathrm{Desc}(r_j)| \quad \text{for all } 1 \le i < j \le |R|.
\]

\paragraph{Reactive $Y$ measurement accounting.}
As discussed in \cref{sec:y_states}, we model $\ket{Y}$ state demand from reactive $Y$ measurements stochastically. To do this, we prepare $\ket{Y}$ states in advance for potential corrective $Y$ measurements and, in the next logical cycle, sample whether a correction is needed. If a prepared $\ket{Y}$ state is not used, it remains in memory until it is consumed by a later gate. In \cref{sec:cost_derivation_reactive_y_methods}, we show in detail how the scheduler tunes the $\ket{Y}$ state production target to balance the storage cost of surplus states against the penalty cost of insufficient states incurred when stochastic demand exceeds supply.


The pseudocode for the Greedy sorting algorithm is then given by~\cref{alg:greedy-block-scheduler}, where we have used the DAG vertex property definitions from~\cref{sec:dag}. For simplicity, we omit dedicated pseudocode for $\ket{Y}$ state usage in \cref{alg:greedy-block-scheduler}, the relevant costs can be thought of as being included in the AV of gates and memory requirements for data qubits.

For a given circuit, the greedy scheduler has a minimum number of logical qubits for which it works, since each logical cycle must provide enough memory for the data qubits. If the memory requirement is too large we may be unable to schedule any more gates, thus producing a~\emph{quantum out-of-memory error} (QOOM).

\subsubsection{Validation of Greedy Block Scheduler}

\noindent By construction the scheduler assigns qubits to one of the five defined roles in each logical cycle, see~\cref{sec:terminology}, such that all qubit counts are nonnegative and their sum equals the total number of logical qubits. To check the correctness of the Greedy Scheduler, we created unit tests that check the scheduler's output on small circuits and edge cases. We also perform some simple checks for real circuit instances:
\begin{itemize}
    \item Check that the total AV of the scheduled gates equals the circuit AV.
    \item Check that all gates are scheduled.
\end{itemize}

After performing these checks, we plotted the per-qubit utilization for the test circuit described in~\cref{sec:test_circuit_description}. The resulting breakdown, shown in~\cref{fig:qubit_usage_per_logical_cycle}, appears reasonable as bridge qubits and stale states constitute only a very small fraction of the total. Moreover, the plot clearly delineates the distinct subcircuit regions of the test circuit, providing additional evidence that the intended gates are being scheduled correctly.

\section{Resource Estimation}
\label{sec:resource_estimation}

\noindent In this section we discuss two specific methods that were used to generate resource estimates: analytic and block-scheduled. Historically, authors have primarily used analytic methods~\cite{Litinski22Active, litinski2023compute, Caesura25} to estimate resources by simply adding up the AV and plugging its value into an equation. Although some optimization was needed, this method was very fast, albeit coarse. This is in contrast to the novel block-scheduled method presented in this paper. The block-scheduled resource estimation method provides a more accurate and optimized resource estimate, but is much slower. We find that using the analytic resource estimation method as an initial guess can greatly cut down on the time required for the block-scheduled method. Later, we will use the results of the block scheduler method to refine previous analytic resource estimation techniques.

\subsection{Analytic Resource Estimation}
\label{subsec:ARE}
\noindent We follow the analytic resource estimation (ARE) methods of~\cite{Caesura25}, where the sizes of the workspace and memory are allowed to differ. Like other ARE methods, all estimates assume access to a decoder than is fast enough to prevent stale state accumulation. The memory consists of data qubits, bridge qubits, and stale states, as in~\cref{fig:qubit_classifications}. To estimate the number of memory qubits we take the maximum number of data qubits used at any point in the circuit, $m_{max}$, and add 20\% to account for bridge qubits and stale states \cite{Litinski22Active}. Thus, we will write
\begin{align}
    m = 1.2 \times m_\textrm{max}
\end{align}
for the analytic resource estimates, where $m$ is the memory size.

After taking the memory into account all remaining qubits are assigned as workspace. We assume that each qubit in the workspace executes one AV block each logical cycle. Thus the number of logical cycles (and therefore the total time of the algorithm) can be derived by counting the AV and dividing it by the workspace size as described in~\cref{subsec:prior_ARE_methods}. After decomposing the test circuit into simple gates using the methods found in~\cref{sec:test_circuit_description}, the AV table (Table~1 in~\cite{litinski2023compute}) was used to sum up the total AV of the computation.

Following this prescription the authors of~\cite{Caesura25} derived the following equation for the time to completion $t$ to execute $V$ AV blocks on a photonic quantum computer with memory size $m$:
\begin{align}
    t 
    &= \frac{Vd^3}{n_{\textrm{IM}} r_\textrm{IM} - \frac{c_\textrm{fiber}}{l_\textrm{delay}}md^2},\label{eq:explicit_comp_time}
\end{align}
where $V$ is the AV of the computation, $d$ is the code distance, $n_\textrm{IM}$ is the number of interleaving modules, $r_\textrm{IM}$ is the rate at which resource states are produced by a single interleaving module, $l_\textrm{delay}$ is the delay length, and $c_\textrm{fiber}$ is the speed of light in fiber. Moreover, the authors of~\cite{Caesura25} were able to derive the maximum possible AV that can be executed by a photonic quantum computer as
\begin{align}
\label{eq:computable_blocks3}
    V_\textrm{max} &\approx \left(1 - m\frac{d^2 c_\textrm{fiber}}{n_{\textrm{IM}} r_\textrm{IM}l_\textrm{delay}}\right) p_{\mathrm{f}} 10^{\frac{\alpha d}{2}},
\end{align}
where $p_\mathrm{f}$ is the total allowed probability of hardware failure and $\alpha$ is a parameter that characterizes the probability that a physical error happens at each spacetime block after quantum error correction is applied. The values chosen for each of these hardware parameters are summarized in~\cref{tab:hardware_parameters}. These parameters were chosen so that the computer is roughly half the physical size of the computer considered in~\cite{Caesura25} to simulate an earlier generation of computer.

Resource estimates are performed by first counting up the total AV in the circuit and finding the smallest integer $d$ such that $V < V_\textrm{max}$. This step ensures that the computation only incurs a hardware error with probability $p_\mathrm{f}$. This step is fast for the ARE case because we can quickly search values of $d$ using~\cref{eq:computable_blocks3}. We will find that in the block scheduler resource estimation this step is not as fast. Once the code distance is known, we can plug it into~\cref{eq:explicit_comp_time} to get the time to completion.

\begin{figure}
    \centering
    \includegraphics[width=1\linewidth]{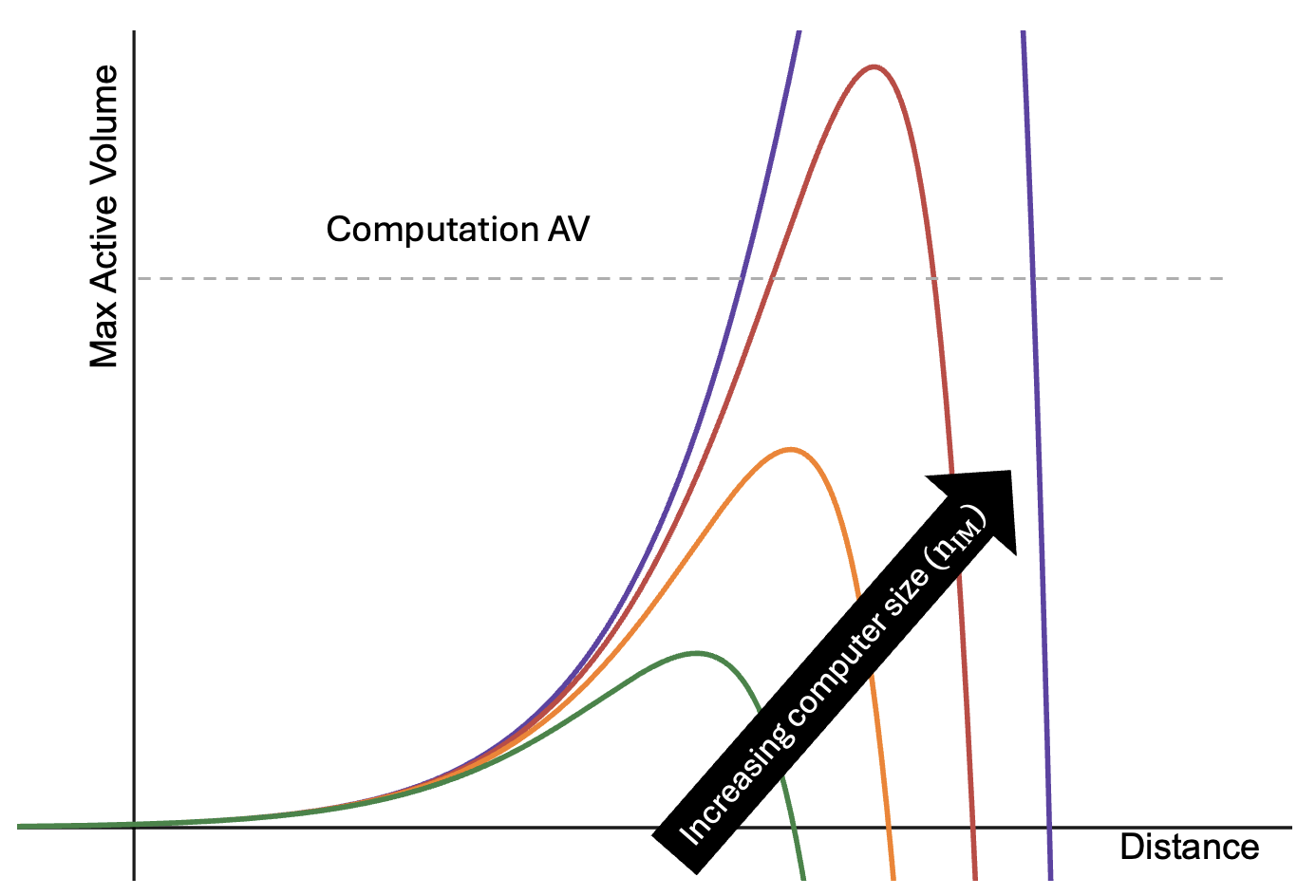}
    \caption{A qualitative plot of~\cref{eq:computable_blocks3} that graphs the maximum AV that can be executed as the size of the computer increases. This is achieved by increasing the number of interleaving modules from the lowest (green) to the highest (purple). Notice that for each computer size, there is an optimal distance at which the largest possible AV can be executed and after that the amount of AV that can be executed drops precipitously. This graph illustrates the intuition that there are some computations that cannot be completed without increasing the size of the computer, even if a large distance is chosen. For a computation of size given by the dashed line, only the 2 largest computer sizes will finish the computation.}
    \label{fig:qualitative_max_av}
\end{figure}

It is possible for the resource estimation to fail if the computer is not large enough for the computation. This can be qualitatively seen in~\cref{fig:qualitative_max_av} where the maximum number of AV blocks peaks at a certain distance and then drops rapidly. If a suitable distance is not found before that drop off occurs, then the resource estimate will fail, and the computer size must be increased in order for the computation to complete.

If the ARE fails to find a functioning code distance for which $V < V_\textrm{max}$, it is still important to get an estimate of the code distance for use in the block scheduler. Therefore, we also calculate the distance which minimizes $V - V_\textrm{max}$. This can then be used as a starting point for the distance search in the block scheduler resource estimation. For a precise description of the ARE see~\cref{alg:ARE}.

\begin{table}[tp]
    \centering
    \footnotesize
    \begin{tabular}{lc}
        \toprule
         Hardware parameter & Value\\
        \midrule
        number of interleaving modules ($n_\textrm{IM}$) & 30\\ 
        resource state production rate ($r_\textrm{IM}$) & $1.6\times10^9\,\mathrm{s}^{-1}$\\
        delay length ($l_\textrm{delay}$) & $2.5\,\mathrm{km}$\\ 
        speed of light in fiber ($c_\textrm{fiber}$) & $\frac{2}{3}c$\\ 
        probability of hardware failure ($p_\mathrm{f}$) & 10\%\\
        hardware error parameter ($\alpha$) & 0.5\\
        logical cycle time ($t_{lc}$) & $3.8 \times 10^{-4}\,\mathrm{s}$\\
        reaction time & $1.5 \times 10^{-5}\,\mathrm{s}$\\
        \bottomrule
    \end{tabular}
    \caption{A summary of the free parameters that were set to create the hardware model used in the example computation. The logical cycle time was calculated with a distance of 30 and is included to give order of magnitude estimates to compare with the reaction time. These parameters were selected to ensure feasibility on an early fault-tolerant photonic quantum computer based on~\cite{Litinski22Active} and are not intended to reflect PsiQuantum’s product roadmap.}
    \label{tab:hardware_parameters}
\end{table}

\subsection{Block-scheduled runtime estimation}
\label{subsec:TSRE}

\noindent Next, we consider the time to completion from the block-scheduled circuit. A precise version of the algorithm can be found in~\cref{alg:TSRE}. From the block scheduler described in~\cref{sec:compiler_description} we can get the number of logical cycles required to complete a computation, $n_{lc}$. We calculate the number of logical cycles by breaking up a representative example of an early fault-tolerant quantum algorithm into the subcircuits mentioned in~\cref{sec:test_circuit_description}
\begin{align}
    n_{lc}  = (2^{n_p}-1) n_{C-U} + n_{QFT},
\end{align}
where $n_{C-U}$ and $n_{QFT}$ are the number of logical cycles needed for the controlled unitary and QFT respectively. $n_p$ is the number of phase qubits used in the computation. $n_{C-U}$ and $n_{QFT}$ are obtained by running the gates described in~\cref{sec:test_circuit_description} through the scheduler described in~\cref{sec:compiler_description} and counting the logical cycles. The factor of $2^{n_p}-1$ comes from the fact that we repeat the controlled U subcircuit $2^{n_p}-1$ times throughout the computation. We refer to these as subcircuit multiplicities.

This method of compilation means that we cannot optimize the compilation routine across subcircuits, which leads to a slight decrease in the efficiency of the scheduler. However, this was necessary for the compilation to finish within a reasonable time frame. It is important to note that $n_{lc}$ depends implicitly on $d$. This is because when we increase the code distance, all other things being equal, we need to decrease the number of qubits in the computer to account for using more resources per qubit. When we decrease the number of qubits in the computer, we have fewer workspace qubits per logical cycle for executing gates, which means the computation progresses slower.

Once we have the total number of logical cycles, we can use the time per logical cycle 
\begin{align}
    t_\textrm{lc} = \frac{l_\textrm{delay} d}{c_\textrm{fiber}}
\end{align}
 to determine the total time to completion of the block scheduled algorithm 
\begin{align}
\label{eq:time_scheduler_time}
    t_\textrm{TS} = n_{lc}(d)\frac{l_\textrm{delay} d}{c_\textrm{fiber}},
\end{align}
where $n_{lc}$ is explicitly written as a function of $d$. However we do not expect that this will work with any distance, we must determine a distance that is sufficient to accommodate the full runtime of the algorithm. This is done by calculating the maximum possible time that a computer can take for given distance by combining~\cref{eq:explicit_comp_time} and~\cref{eq:computable_blocks3} to get
\begin{align}
\label{eq:max_time}
    t_\textrm{max} &= \frac{d^3}{n_{\textrm{IM}} r_\textrm{IM}}\; p_{\mathrm{f}}\, 10^{\frac{\alpha d}{2}}
\end{align}
and comparing it to $t_\textrm{TS}$. The smallest distance for which $t_\textrm{TS} < t_\textrm{max}$ is true will minimize the runtime of the algorithm~\footnote{There may be pathological cases where the greedy scheduler may perform better at larger distances, but these cases are contrived and unlikely to occur in a real-world computation.}.

One may note that the optimization loop described in the previous paragraph requires the entire compilation pipeline to be rerun at each iteration. This would usually mean that the optimization loop is very slow and would take a long time to find the minimum distance.
However, we find that in practice the distance used for ARE is always larger than the block scheduler distance in non-reaction limited regimes (where the ARE model is valid). We therefore start at the ARE code distance and iteratively decrease it until either $t_\textrm{TS} < t_\textrm{max}$ or the compiler runs out of space to schedule gates and fails with an out-of-memory error. In practice this requires at most three compilation passes in the successful case and around 5 or so in the failure case. Along with caching the gate DAGs for reuse (since they are independent of code distance), this reduces the compilation time of the test circuit to a few minutes.

\begin{algobox}[\textsc{Analytic Resource Estimate (ARE)}]
\label{alg:ARE}
\begin{algorithmic}[1]
\Require Operation DAG $\mathcal{R}$, candidate distances $\mathcal{D}=\{1,2,\dots,d_{\max}\}$, limit $V_\textrm{max}$
\Ensure A time to completion $t$ and feasible code distance $d^\star$ if one exists; otherwise report failure and the distance $d_{\Delta\textrm{min}}$ which came closest to succeeding.

\State $V \gets 0$ \Comment{Sum AV from each gate}
\ForAll{$v \in \mathcal{R}$}
  \State $V \gets V + \mathrm{AV}(v)$
\EndFor

\State $\Delta V_\textrm{min} \gets \infty$, $d_{\Delta\textrm{min}} \gets -1$

\For{$d \gets 1$ to $d_{\max}$}
  \State Compute $V_\textrm{max}(d)$ using~\cref{eq:computable_blocks3}.
  \If{$V < V_\textrm{max}(d)$} \Comment{Return smallest viable distance.}
    \State Calculate $t$ using~\cref{eq:explicit_comp_time} for distance $d$.
    \State \Return $(\textbf{ok}, d, t)$
  \EndIf
  \State $\Delta V \gets |V - V_\textrm{max}(d)|$ \Comment{Track minimum difference in volume}
  \If{$\Delta V < \Delta V_\textrm{min}$}
    \State $d_{\Delta\textrm{min}} \gets d$
    \State $\Delta V_\textrm{min} \gets \Delta V$
  \EndIf
\EndFor

\State \Return $(\textbf{FAILURE}, d_{\Delta\textrm{min}}, \infty)$
\end{algorithmic}
\end{algobox}

\begin{algobox}[\textsc{Block Scheduler\\ Resource Estimate}]
\label{alg:TSRE}
\begin{algorithmic}[1]
\Require Number of subcircuits $n_{sc}$, Candidate distances $\mathcal{D}=\{1,2,\dots,d_{\max}\}$, gate DAGs for each subcircuit $\mathscr{R} = \{\mathcal{R}_0, \mathcal{R}_1, ..., \mathcal{R}_{n_{sc}}\}$, Subcircuit multiplicities $\mathcal{M} = \{m_0, m_1, ...., m_{n_{sc}}\}$
\Ensure A time to completion $t$ if a feasible code distance $d^\star$ if one exists; otherwise report failure
\State $d \gets $ output from~\cref{alg:ARE}.
\State $t_\textrm{min} \gets \infty$
\State $\mathcal{D}_\textrm{seen} \gets \emptyset$
\While{$d \in \mathcal{D}$ and $d \notin \mathcal{D}_\textrm{seen}$}
  \State $\mathcal{D}_\textrm{seen} \gets \mathcal{D}_\textrm{seen} \cup \{d\}$
  \State $n_{lc} \gets 0$
  \For{$g \gets 1$ to $n_{sc}$}
    \State Use~\cref{alg:greedy-block-scheduler} on $\mathcal{R}_g$ to calculate $n_{lc}^g$.
    \If{\cref{alg:greedy-block-scheduler} terminates due to QOOM}
      \State $d \gets d - 1$ \Comment{Needs more qubits.}
      \State Start next while iteration
    \EndIf
    \State $n_{lc} \gets n_{lc} + m_g \cdot n_{lc}^{g}$
  \EndFor
  \State Use~\cref{eq:time_scheduler_time} to calculate $t$
  \State Use~\cref{eq:max_time} to get $t_\textrm{max}$
  \If{$t > t_\textrm{max}$}
    \State $d \gets d + 1$ \Comment{Needs stronger error correction.}
  \Else
    \State $t_\textrm{min} \gets \min(t, t_\textrm{min})$
  \EndIf
\EndWhile
\If{$t_\textrm{min} < \infty$}
  \State \Return $t_\textrm{min}$
\Else
  \State \Return \textbf{FAILURE}
\EndIf
\end{algorithmic}
\end{algobox}

\section{Results}
\label{sec:results}

\noindent We evaluate the Greedy Block Scheduler using qubitized Fermi--Hubbard simulation circuits in an Active Volume architecture with the properties described in~\cref{subsec:ARE}. The central result of this section is that explicit scheduling reveals a workspace-dependent transition to a reaction-dominant regime. Below this transition, additional logical qubits reduce runtime by increasing the workspace available for logic execution. Above it, reaction latency limits the effective workspace and substantially reduces the runtime benefit obtained from adding more qubits.

We first define the reaction-dominant regime and identify its onset using reaction-depth and workspace statistics. We then show how this transition changes runtime scaling relative to analytic resource estimates that neglect reaction latency. Next, we use representative non-reaction-dominant instances to test the assumptions made in~\cite{Litinski22Active}, quantify bridge, stale state, and routing overheads. Finally, we benchmark the scalability and near-optimality of the scheduler. The Fermi--Hubbard instances considered here are intended as demonstration workloads for studying execution effects in an Active Volume architecture, rather than as state-of-the-art resource estimates for quantum simulation.

\subsection{A workspace-dependent reaction-limited regime}
\label{subsec:reaction_dominant_regime}

\noindent We begin by studying when reaction latency becomes an execution bottleneck. For this study, we schedule the full qubitized Fermi--Hubbard circuit rather than scheduling the unitary and QFT subcircuits independently, since reaction dependencies can extend across the artificial boundaries introduced by circuit decomposition. To make this full-circuit sweep tractable at large computer sizes, we reduce the number of bits in the phase register by a factor of two. The numerical thresholds reported in this subsection should therefore be interpreted as thresholds for this reduced-precision full-circuit workload. We use this data to identify the conditions that start the reaction-dominant transition, rather than to give a final full-precision resource estimate.

Let $R_{\mathrm{LC}}=25$ denote the number of reaction layers that can be completed within one logical cycle. For each logical cycle $\ell$, let $r_\ell^{\max}$ be the largest reaction layer count of any stale state in that cycle. We call a logical cycle \emph{reaction-limited} when
\begin{equation}
r_{\ell-1}^{\max} > R_{\mathrm{LC}}.
\end{equation}
Since every stale state in our model occupies memory for at least one logical cycle, this criterion identifies cycles in which at least one stale state must be retained beyond that minimum lifetime. We define the \emph{first reaction-limited point} as the smallest computer size for which any logical cycle satisfies this condition. We define the onset of the \emph{reaction-dominant regime} operationally as the point at which more than half of logical cycles are reaction-limited. The latter definition is more relevant for resource estimation, because isolated reaction-limited cycles may have little effect on the total runtime, whereas reaction-limited cycles appearing throughout the computation change the effective scaling of the schedule.

\begin{figure}[!tbp]
    \centering
    \includegraphics[width=\linewidth]{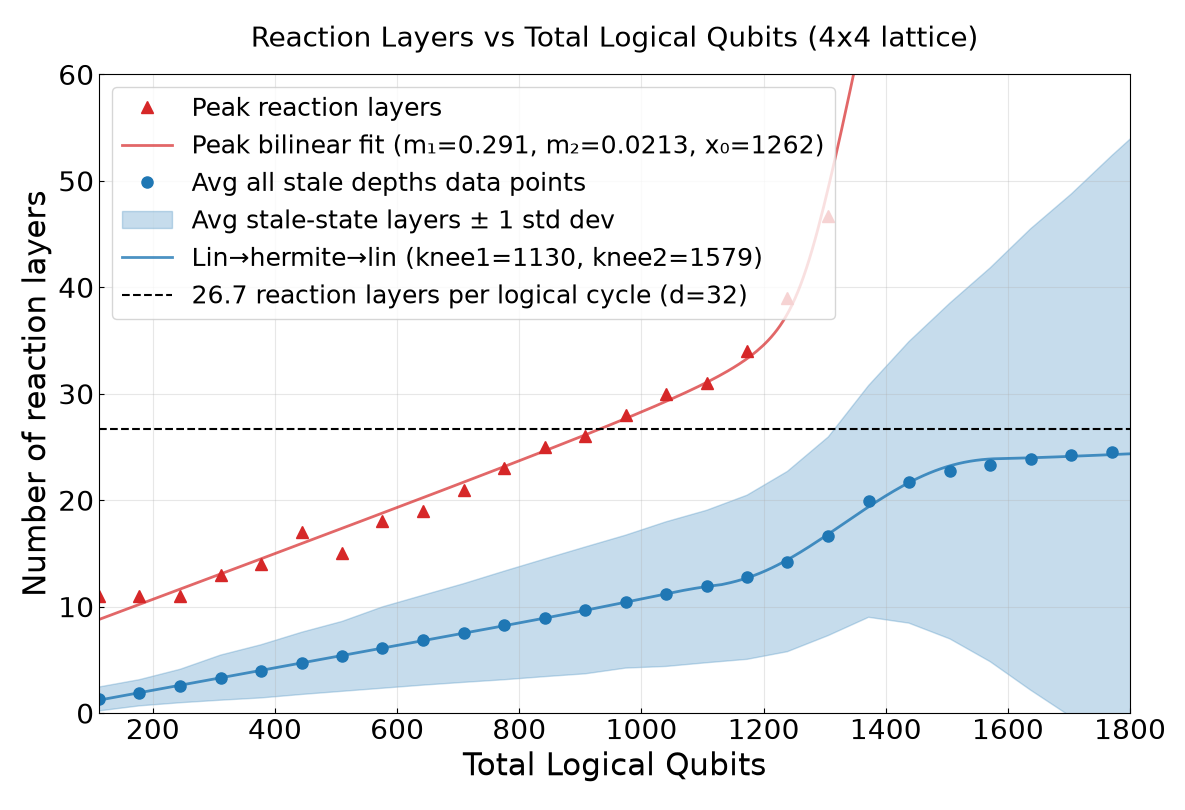}
    \caption{Peak (red) and mean (blue) reaction layers across all logical cycles for a block-scheduled $4\times4$ qubitized Fermi--Hubbard circuit, plotted as a function of logical qubit count. The shaded blue band shows one standard deviation about the mean stale state layer count. The mean reaction layer count grows approximately linearly up to about 1130 qubits, where it undergoes a Hermite transition to a shallower linear trend. The peak reaction layer count grows approximately linearly up to about 1200 qubits, beyond which it transitions to a steeper linear trend.}
    \label{fig:reaction_depth_vs_computer_size}
\end{figure}

In~\cref{fig:reaction_depth_vs_computer_size}, the first stale states with reaction depth comparable to one logical cycle appear at approximately 950 logical qubits. This marks the first reaction-limited point, but not yet the reaction-dominant regime: only a small fraction of cycles are affected. As the computer size is increased further, the fraction of logical cycles containing stale states with more than 26 reaction layers rises rapidly. At approximately 1200 logical qubits, about 50\% of logical cycles are reaction-limited. We therefore identify this point as the onset of significant reaction-limited behaviour, and use it as the operational onset of the reaction-dominant regime.

The same transition is visible in the average stale state reaction depth. We fit the average stale state layer data using two linear regimes connected by a Hermite-smoothed transition. The first knee occurs at approximately 1130 logical qubits, close to the 50\% reaction-limited-cycle point. The second knee occurs at approximately 1579 logical qubits, after which the fitted slope is much shallower. Around the middle of this transition, at approximately 1320 logical qubits, the one-standard-deviation envelope reaches the one-logical-cycle threshold of 26.7 reaction layers. These features show that the transition is smooth: rare long reaction chains appear first, then reaction-limited cycles become common, and finally the average behavior of the computation is modified.

\begin{figure}[!tbp]
    \centering
    \includegraphics[width=\linewidth]{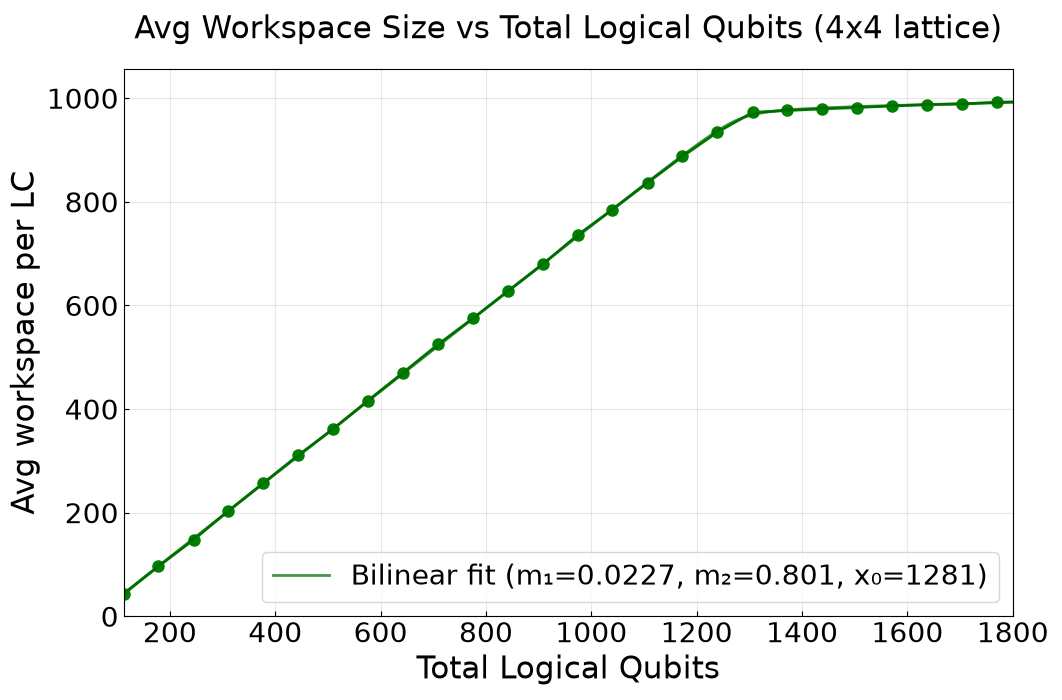}
    \caption{The average workspace size across all logical cycles for a block-scheduled $4\times4$ qubitized Fermi--Hubbard circuit, plotted as a function of logical qubit count. The workspace grows linearly up to approximately 1300 qubits, after which it rapidly transitions to a line with a much shallower gradient. The transition point aligns with the onset of the reaction dominant regime, see~\cref{fig:reaction_depth_vs_computer_size}.}
    \label{fig:workspace_vs_computer_size}
\end{figure}

This interpretation is supported by the average workspace data in~\cref{fig:workspace_vs_computer_size}. At small and intermediate computer sizes, the average workspace grows approximately linearly with the total number of logical qubits. This is the expected workspace-limited regime: additional qubits can be converted directly into additional logic execution. Around 1300 logical qubits, however, the average workspace switches to a shallower linear trend. This indicates a feedback mechanism. Increasing the workspace allows more operations to be executed per logical cycle, which produces more stale states. Once reaction latency prevents those stale states from being cleared within a single logical cycle, they persist in memory and accumulate, reducing the workspace available in subsequent cycles. The computation then approaches an equilibrium between logic execution and the stale states that this execution produces.

We therefore identify the reaction-dominant regime not simply with the existence of any stale state deeper than one logical cycle, but with the onset of this feedback mechanism. In this regime, adding more logical qubits gives substantially worse runtime returns, because the effective workspace is increasingly limited by unresolved stale states rather than by the nominal computer size.

\subsection{Verifying and refining ARE assumptions}
\label{sec:scheduler_vs_are}

\noindent Here we test the analytic resource estimator (ARE) model used in~\cite{Caesura25}, which follows the Active Volume estimate of~\cite{Litinski22Active}, but replaces the fixed 50/50 memory--workspace split with a memory allocation based on the maximum number of data qubits used at any point in the circuit (qubit highwater). This gives a more circuit-specific estimate of the memory requirement, but it still makes several assumptions that are not resolved at the logical cycle level. In particular, ARE assumes that workspace can be packed efficiently, that the qubit highwater is a good proxy for average memory requirements, that bridge and stale state overheads are accounted for by adding a 20\% overhead to the memory allocation, and that reaction latency does not limit the computation.

Having identified the onset of reaction-dominant behavior in~\cref{subsec:reaction_dominant_regime}, we now restrict attention to the workspace-limited regime in which the ARE model is valid. For the full-circuit workload studied above, the first stale states lasting longer than one logical cycle appear at approximately 950 logical qubits, while reaction-limited cycles become common at approximately 1200 logical qubits. Analytic estimates like ARE that ignore reaction latency should not be extrapolated beyond this regime unless the effective reaction time can be reduced enough that all stale states are still cleared within one logical cycle. In this subsection, we therefore compare the scheduler with ARE at a smaller computer size, where the peak reaction layers is only 7, well below the one-logical-cycle threshold $R_{\mathrm{LC}}=26.7$.

For this comparison, we use a component-wise scheduling procedure that makes the full-precision phase register simulation tractable. We split it into the controlled-unitary and inverse-QFT components. Each component is scheduled once and then multiplied by the number of times it appears in the full circuit. If component $i$ has schedule length $L_i$ and appears $m_i$ times, the total schedule length is approximated as
\begin{equation}
L_{\mathrm{parts}}=\sum_i m_i L_i .
\label{eq:parts_schedule_length}
\end{equation}
The main approximation of this procedure is that the final logical cycle of each component may be artificially under-filled: in a full-circuit schedule, operations from the next component could sometimes be scheduled into this otherwise sparse final cycle. Secondly, stale states are lost at the boundary, which under estimates the effect of reaction latency. We therefore enforce the boundary-cycle fraction
\begin{equation}
\label{eq:boundary_threshold}
\epsilon_{\mathrm{bdry}}
=
\frac{\sum_i m_i}{\sum_i m_i L_i}
\end{equation}
to be less than $3.80\%$. For example, if a component with 100 logical cycles is used 50 times, then only 50 out of the resulting 5000 logical cycles are component-final cycles, giving $\epsilon_{\mathrm{bdry}}=1\%$. $3.8\%$ is obtained empirically by determining the $\epsilon_{\mathrm{bdry}}$ that gives us a component-wise (parts) runtime within $5\%$ of the full-DAG version, see~\cref{sec:validity_of_parts}. Component-wise scheduling is therefore appropriate for the low-qubit-count comparison in this subsection, but not for the large-workspace reaction-dominant study above.

Using this component-wise construction, we compare the scheduler and ARE for a $4\times4$ Fermi--Hubbard instance using equivalent photonic resources. The results are shown in~\cref{tab:prev_compilation_comparison}. This comparison isolates differences in packing, memory allocation, and bridge/stale state overheads.

\begin{table}
\centering
\small
\begin{tabular}{@{}lcc@{}}
\toprule
Attribute & \makecell{Scheduler\\(this paper)} & \makecell{ARE~\cite{Caesura25}} \\
\midrule
Logical qubits & 195 & 183 \\
Code distance & 32 & 33 \\
Time to completion (s) & 853 & 1400 \\
\addlinespace
\multicolumn{3}{@{}l}{\emph{Average fraction of total qubits}} \\
\quad Memory & 28.8\% & 51.9\% \\
\quad Bridge + stale state & 12.0\% & 10.4\% \\
\quad Unused & 5.44\% & 0\% \\
\quad Routed & $\approx 25\%$ & N/A \\
\addlinespace
Peak reaction layers & 7 & N/A \\
\bottomrule
\end{tabular}
\caption{Metrics for simulating a $4\times4$ Fermi--Hubbard lattice obtained using the block scheduler and ARE using equivalent photonic resources. Here, $\eta=-1.8$ is the $\ket{Y}$ state production bias parameter. The average fraction of qubits required for memory in the scheduler model is $1.80\times$ less than the ARE model, enabling a larger workspace size which substantially reduces the runtime and lowers the code distance. A lower code distance allows for more logical qubits, which are added to the workspace thus decreasing runtime further. The scheduler predicts a $1.64\times$ speedup in computation time relative to the ARE model. However, see that the ARE slightly underestimates bridging and stale state requirements.}
\label{tab:prev_compilation_comparison}
\end{table}

The scheduled computation leaves $5.44\%$ of logical qubits unused on average. This indicates that the greedy scheduler is able to allocate most available qubits to data storage, workspace, bridge qubits, or stale states, but also that the zero-unused-qubit approximation made in ARE is optimistic. In this instance, the resulting error is a few percent of the logical qubit budget, rather than the dominant source of discrepancy.

The largest discrepancy remains the data qubit estimate. ARE allocates memory using the qubit high-water mark, whereas the scheduler tracks the data qubit requirement in each logical cycle. For this instance, ARE assigns $51.9\%$ of logical qubits to memory, while the scheduler requires only $28.8\%$ on average. Thus, the high-water mark remains a conservative proxy for memory usage, overestimating the average memory occupancy by a factor of $1.80$. This distinction is important because qubits not occupied by memory can instead be used as workspace. The additional workspace allows more operations to be executed per logical cycle, reducing the total number of logical cycles required. This shorter runtime reduces the spacetime volume associated with stored data qubits and, in this instance, is sufficient to lower the required code distance, which frees up more qubits and reduces the runtime further.

In contrast to the memory estimate, ARE slightly underestimates the bridge and stale state (BSS) qubit overhead. ARE assigns $10.4\%$ of logical qubits to the BSS overhead, while the scheduler requires $12.0\%$ on average. We combine bridge qubits and stale states in this comparison only because ARE treats them as a single overhead term. In the scheduler, they are distinct resources: bridge qubits connect operations within a logical cycle, whereas stale states arise from measurements whose bases depend on previous outcomes.

\subsubsection{Bridge and stale-state overhead}
\label{subsubsec:bss_overhead}

\begin{figure}
\centering
\includegraphics[width=\linewidth]{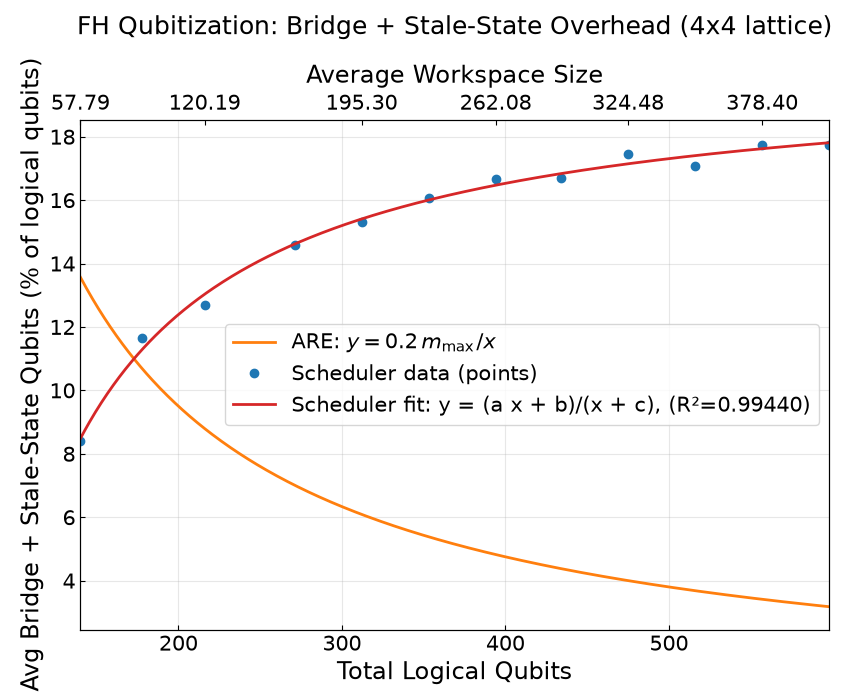}
\caption{Mean bridge and stale state overhead for a scheduled $4\times4$ Fermi--Hubbard circuit as the number of logical qubits is varied. Since the circuit is fixed, the ARE model predicts a decreasing overhead fraction, $f_{\mathrm{BSS}}^{\mathrm{ARE}}(x)=0.2m_{\max}/x$, with $m_{\max}=95$. In contrast, the scheduler shows that bridge and stale-state overhead grows with available workspace. The scheduler data are fit by $f_{\mathrm{BSS}}^{\mathrm{sched}}(x)=(a x+b)/(x+c)$, with $a=0.2027$, $b=-18.13$, and $c=19.12$, giving $R^2=0.9944$. The two curves intersect at approximately 180 logical qubits.}
\label{fig:bridging_and_stale_stale_dependency_comparison}
\end{figure}

\noindent To refine the bridge and stale state (BSS) qubit overhead model, we sweep the number of logical qubits and compare the scheduler data with the ARE assumption. In~\cref{fig:bridging_and_stale_stale_dependency_comparison}, we plot the mean BSS overhead as a fraction of the total logical qubit count. For the $4\times4$ Fermi--Hubbard circuit, the maximum data qubit requirement is $m_{\max}=95$. The ARE model from \cite{Caesura25} estimates the BSS overhead as
\begin{equation}
f_{\mathrm{BSS}}^{\mathrm{ARE}}(x)
=
\frac{0.2 \cdot m_{\max}}{x},
\end{equation}
where $x$ is the total number of logical qubits, see~\cref{subsec:ARE}. This expression implements the assumption that BSS overheads are equal to $20\%$ of the memory allocation.

The scheduler data shows that this memory-based scaling has the wrong dependence on computer size. Since the circuit is fixed, $m_{\max}$ is constant, so the ARE bridge/stale state fraction decreases as $1/x$. By contrast, increasing the total number of logical qubits increases the available workspace. A larger workspace allows more operations to be executed in the same logical cycle, which increases both the number of bridge qubits required to connect operations and the number of stale states produced per cycle. The scheduler therefore predicts that BSS overhead grows with workspace, inverting the trend predicted by ARE.

For the $4\times4$ instance, the scheduler data is well fit by the rational function
\begin{equation}
f_{\mathrm{BSS}}^{\mathrm{sched}}(x)
=
\frac{a x+b}{x+c},
\end{equation}
with $a=0.2027$, $b=-18.13$, and $c=19.12$, obtaining a coefficient of determination of $R^2=0.9944$. The scheduler and ARE curves intersect at approximately 180 logical qubits. Below this point, ARE overestimates the bridge/stale-state overhead; above it, ARE underestimates the overhead. At the comparison point in~\cref{tab:prev_compilation_comparison}, this gives a scheduler bridge/stale-state overhead of $12.0\%$, compared with the ARE value of $10.4\%$.

The asymptotic behavior of the scheduler is given by the limit $x\rightarrow\infty$,
\begin{equation}
    \lim_{x\rightarrow\infty}
    f_{\mathrm{BSS}}^{\mathrm{sched}}(x)
    =
    a
    \approx
    20.3\%.
\end{equation}
The scheduler approaches a constant, which is expected as the large-workspace limit corresponds to a single logical cycle regime, where all gates are executed in parallel and bridging is maximized. This value is numerically close to the $20\%$ factor used in~\cite{Litinski22Active}, but this agreement should be viewed as coincidental rather than as validation.

The fitted coefficients should not be interpreted as universal constants. Rather, they show that BBS overhead can be modeled accurately by average workspace size. A more general ARE model should therefore use the rational function form based on the average workspace size, not the memory allocation. Determining a circuit- and workspace-dependent correction across a broader family of algorithms is left to future work.

\subsection{Qubit Routing}
\label{subsubsec:qubit_routing}

\noindent The qubit-routing model of~\cite{Litinski22Active} provides plausible evidence that routing can be completed within the approximately 30 code cycles available during lattice surgery. In this model, qubits participating in lattice-surgery operations remain fixed while other qubits are routed around them. However, the qubit layout required for the next logical cycle cannot always be established within the current logical cycle. Consequently, additional code cycles may be required to complete routing before the next logical cycle can begin. We distinguish between two routing paradigms:

\begin{enumerate}
    \item \textit{Intra-cycle routing:} Routing performed during a logical cycle, subject to constraints imposed by occupied workspace qubits. If all routing can be completed intra-cycle, there is no idling between logical cycles.
    \item \textit{Inter-cycle routing:} Routing performed between logical cycles, during which logical computation is suspended. During inter-cycle routing, all qubits can be routed as workspace qubits have been released.
\end{enumerate}

We use the block scheduler to investigate two cases that require inter-cycle routing:
\begin{enumerate}
    \item \textbf{Accessed data qubit routing.}
    A data qubit participating in an operation cannot be routed during that logical cycle. If it is required at a different location in the following cycle, additional routing may be necessary.

    \item \textbf{Large workspace routing.}
    The workspace layout required by one logical cycle may not be suitable for the operations scheduled in the next. Rearranging memory qubits during a logical cycle becomes increasingly difficult when a large fraction of the computer consists of workspace. Workspace is fixed in place and so limits available routing channels.
\end{enumerate}

We show that the first effect can be accommodated with negligible additional overhead for our test circuit. The second is more difficult to resolve without specifying a detailed physical layout and routing algorithm, so we instead use the scheduler to estimate a conservative upper bound on the required rearrangement. Our results suggest that, although data qubit routing can be handled with little additional overhead in the EFTQC regime considered here, workspace placement and reconfiguration may become the more significant routing challenge as the workspace fraction increases.

\subsubsection{Accessed Data Qubit Routing}

\noindent
Using the block scheduler, we identify operations whose input data qubits participated in different operations during the preceding logical cycle. These are the cases in which additional routing may be required. Data qubits produced by the same preceding operation are already close together, while qubits that were idle in the preceding cycle can be routed toward their next locations during that cycle.

When additional routing is required, the delay can be avoided using the bridge qubit construction of~\cref{fig:bridge_qubits}. Bridge qubits are routed during logical cycle $\ell$, and Bell measurements at the end of the cycle teleport the relevant outputs to locations prepared for their next scheduled operations. Accessed data qubit routing can therefore be removed from the critical path at the cost of a modest increase in bridge qubit demand.

For the test circuit, the block scheduler predicts that this procedure requires an average of only 1.36 additional bridge qubits per logical cycle. This demand can almost always be accommodated using otherwise unused qubits, so accessed data qubit routing contributes negligible additional memory overhead for this circuit. We expect similarly small overheads for other EFTQC computations, since relatively few data qubits participate in different operations in consecutive logical cycles. Future versions of the scheduler could also minimize this additional bridging cost directly.

\subsubsection{Large Workspace Routing}

\noindent
Workspace qubits participating in lattice surgery cannot be moved during a logical cycle. When workspace occupies most of the machine, this can restrict the routing channels available to data qubits. Successive cycles may also require different workspace arrangements to accommodate different sized operations, requiring rearrangement before the next cycle can begin.

To estimate the scale of this effect, we consider the deliberately conservative case in which neither data nor workspace qubits are rearranged until the current logical cycle has completed. We compare a worst-case model, in which as few workspace qubits as possible are already suitably positioned, with an average-case model, in which workspace qubits are assumed to be uniformly distributed throughout the computer. The models, including bridge qubits, stale states, and unused qubits, are defined in~\cref{sec:qubit_routing_models}.

\begin{figure}[!tbp]
    \centering
    \includegraphics[width=\linewidth]{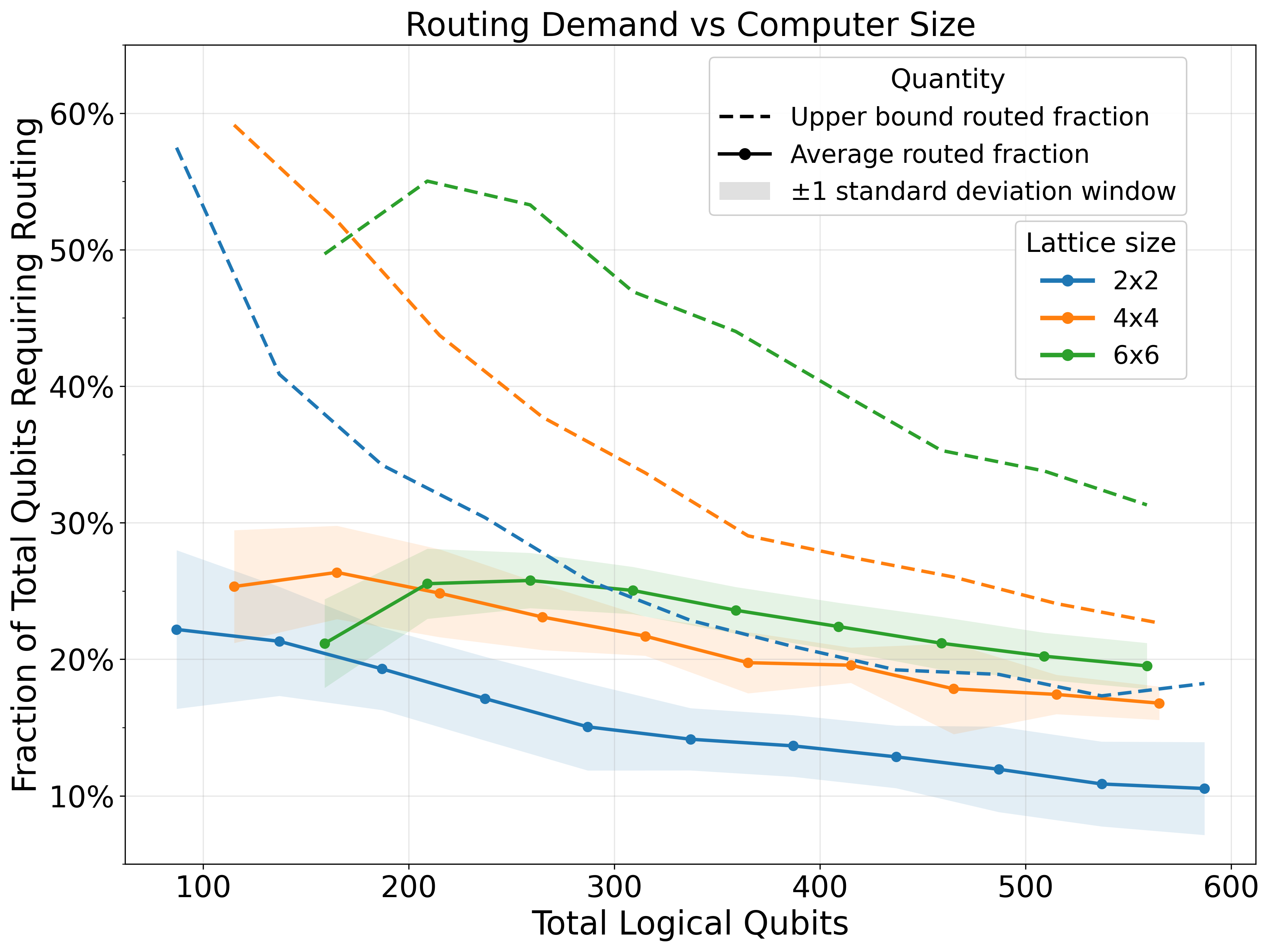}
    \caption{Estimated routing demand for different lattice sizes and machine sizes. For each computation, we evaluate the fraction of qubits that must be routed in each logical cycle. We report the maximum routed fraction over all logical cycles under the worst-case model, as well as the mean routed fraction and standard deviation over logical cycles under the average-case model. The former indicates when routing may begin to affect individual logical cycles, while the latter characterizes the typical routing demand and its variation across the computation.}
    \label{fig:qubit_routing_fraction}
\end{figure}

In~\cref{fig:qubit_routing_fraction}, the worst-case routed fraction reaches at most approximately 50\%, while the average-case routed fraction reaches approximately 30\%. In both models, the routed fraction decreases as computer size increases. This is because larger computers devote a greater fraction of their qubits to workspace, increasing the probability that workspace qubits are already available at the required locations. Despite the decreasing fraction of qubits requiring inter-cycle routing, intra-cycle routing may still present a challenge on larger machines. In Appendix~\ref{sec:routing_runtime_implications}, we discuss possible spacetime tradeoffs that may alleviate this challenge.

\subsection{Scheduler benchmarks}
\label{sec:scheduler_benchmarks}
\noindent To assess the performance of our scheduler, we conducted a series of benchmarks. Comparison with a mixed-integer linear programming (MILP) solver, together with runtime measurements on the test circuit, provides evidence for both near-optimality and scalability. Both sets of results are presented in this subsection.

\begin{figure}
    \centering
    \includegraphics[width=\linewidth]{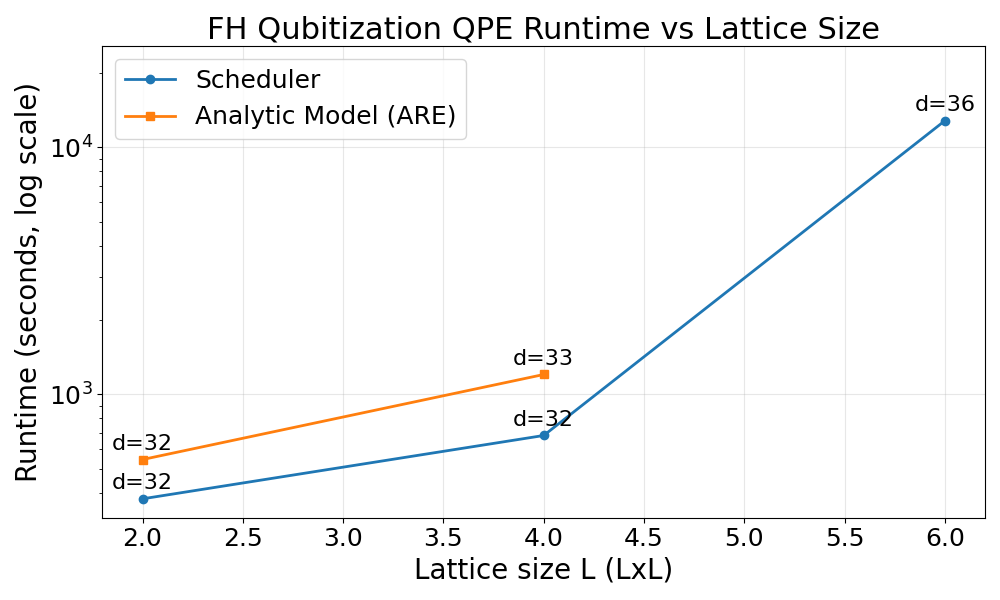}
    \caption{Time to completion for simulations of a Fermi–Hubbard lattice under the analytic and block scheduler models for different lattice sizes. The code distances for each instance are annotated next to the corresponding points. In every instance, the scheduler model predicts faster runtimes than the analytic model. In the $4\times 4$ case, it also enabled a reduced code distance, which reduced the runtime further. This corresponds to speedups of $1.48\times$ for $2\times2$ and $1.76\times$ for $4\times4$. For the $6\times 6$ case, only the scheduler model completed the task; the analytic model’s required code distance left insufficient qubits for data qubit storage.}
    \label{fig:time_comparison_plot}
\end{figure}

\begin{figure}
    \centering
    \includegraphics[width=\linewidth]{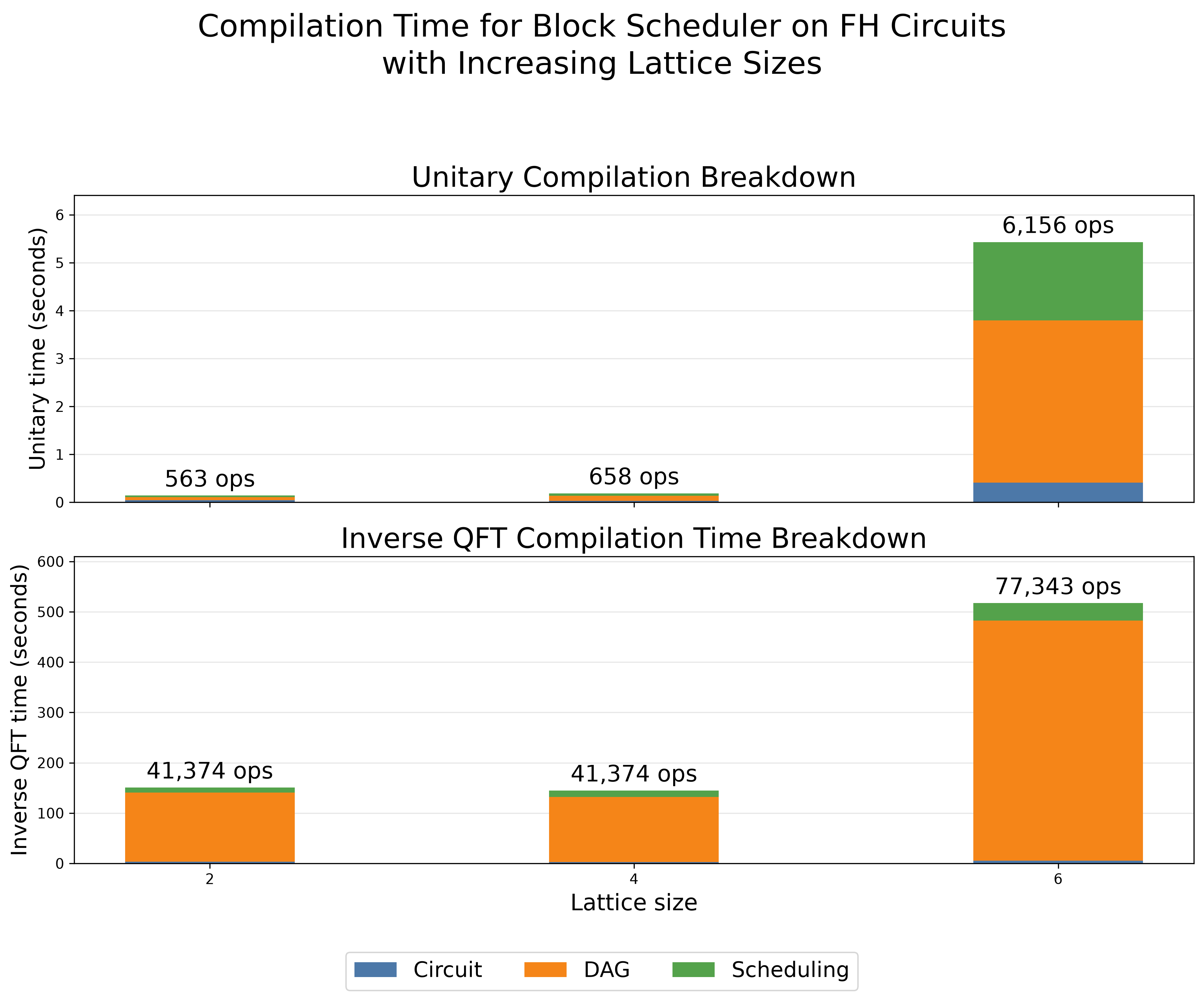}
    \caption{Here we show the time taken by the Block Scheduler to compile the test circuit with different lattice sizes. Within each bar one can see the contribution from the circuit creation, DAG creation, and gate scheduling sections. The vast majority of the time was taken up by the QFT$^\dagger$ section of the compilation. Lattices sizes 2 and 4 have the same phase register size (see~\cref{tab:circuit_parameters}) and so implement the same QFT$^\dagger$. Therefore, any small variations in their compilation time for the QFT$^\dagger$ is due to randomness in the rotation synthesis method and the stochastic simulation of reactive measurements. Lattice size 6 has a larger phase register, giving it a larger QFT$^\dagger$. In all cases, the largest contributor to the compilation time is the DAG creation process. This could likely be sped up with a direct implementation in a lower-level language like C++ or Rust rather than relying on the igraph Python package.}
    \label{fig:compilation_time_breakdown}
\end{figure}

\begin{figure*}[tp]
    \centering
    \includegraphics[width=\textwidth]{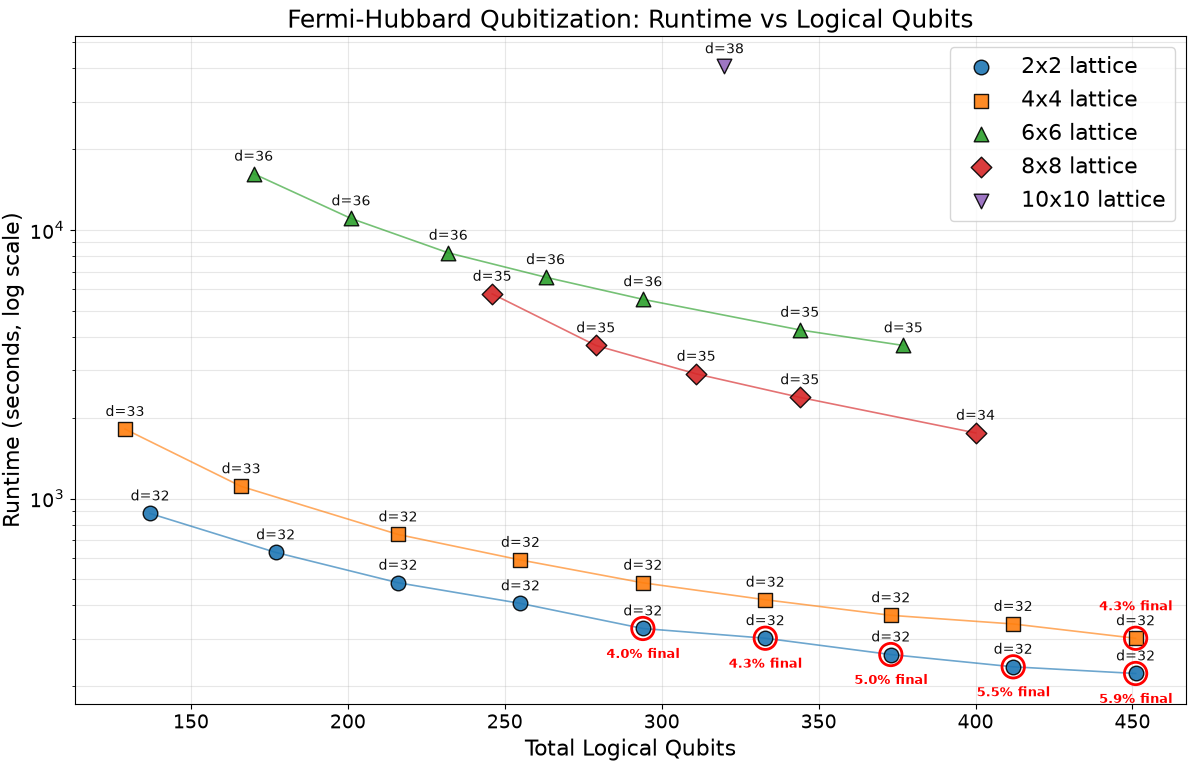}
    \caption{Computation time versus computer size (total logical qubits) for various qubitizes Fermi--Hubbard QPE lattice sizes scheduled in parts by the block scheduler. The qubit count was adjusted by changing the number of interleaving modules $n_\textrm{IM}$. For each computer size, we simulate up to five lattice sizes (see legend). For each lattice, we select the minimum code distance, $d$, that achieves a hardware success probability of 10\% and thereby minimizes runtime. Larger lattices increase the memory required for data qubits, and simulations may fail on smaller machines with a QOOM error if the scheduler workspace is too small to execute any gates. Consequently, the first successful data points for the $8\times8$ and $10\times10$ lattices occur at approximately 245 and 320 logical qubits, respectively. Due to circuit optimizations related to lattice sizes that are powers of 2 the $8\times8$ has less AV and a faster runtime than $6\times6$. The data points circled in red surpass our boundary threshold \cref{eq:boundary_threshold} of $3.80\%$ and so may overestimate runtimes by slightly greater than $5\%$ of a full circuit schedule, see~\cref{sec:validity_of_parts}.}
    \label{fig:rt_against_logical_qubits}
\end{figure*}

\subsubsection{Evidence for optimality}

\noindent To show the near-optimality of the greedy scheduler, we formulated the scheduling problem as a MILP problem and solved it using the CBC solver via the PuLP Python package~\cite{pulp}. Here we focus on the results; see~\cref{sec:MILP_specs} for implementation details.

Since MILP solvers are exact optimization methods, they can in principle certify optimality, although reaching and proving optimality can become computationally demanding on large instances~\cite{Wolsey2020,Kemminer2024}. We used the solver in two complementary ways, each aimed at identifying an ordering of gates that minimizes the number of logical cycles:
\begin{enumerate}
\item \textbf{Optimality certification on a small instance.} To obtain a clear case where the scheduler achieves optimality, we applied the MILP formulation to a smaller 307-vertex QROM DAG (e.g., 1.1.3 of \cref{fig:full_test_circuit}). In this case, the solver was able to prove that the greedy schedule is optimal.
\item \textbf{Warm-started search on a larger instance.} To test whether the MILP solver could outperform the greedy schedule on a larger instance, where greater scheduling flexibility might be available, we considered a 658-vertex DAG representing the unitary gate of a $4\times4$ FH circuit (i.e.,\ 1.1 of \cref{fig:full_test_circuit}). We then reformulated the MILP as a feasibility problem: for a fixed number of logical cycles, $k$, the solver accepted any feasible schedule. Starting from the greedy solution, we decreased $k$ by one at each step and checked for feasibility, using the greedy schedule as a warm start. Even after 12 hours, the solver was unable to improve upon the greedy result by a single logical cycle.
\end{enumerate}
Taken together, these results provide an exact proof of optimality on a smaller instance and strong evidence of near-optimality on the larger instance.

These optimality results are perhaps unsurprising, since the MILP solver has limited freedom to improve the schedule beyond reducing bridge-qubit overhead and minimizing unused qubits. Since both of these costs are already small (see~\cref{tab:prev_compilation_comparison}), there is relatively little room for improvement.

\subsubsection{Scalability Benchmarks}

\noindent Next, in~\cref{fig:time_comparison_plot} we directly compared the time to completion of the quantum computer using both ARE and block scheduling methods for different lattice sizes. We found that in all cases the runtime for the scheduler was less than predicted by ARE and that this difference grows with the lattice size. In the $4\times4$ case, the scheduler achieved a runtime that is fast enough to lower the code distance. This allowed the computer to hold more logical qubits for the same number of resource state generators (physical qubits) and reduced the runtime further. For the case of a $6\times6$ lattice, no suitable distance could be found for the ARE method, while the scheduler yielded a runtime of roughly 3.5 hours. These results serve to demonstrate that the block scheduler provides a more accurate and more optimized picture of the computation than ARE.

In addition to fast QPU times, we also demonstrate in~\cref{fig:compilation_time_breakdown} that this compiler finishes in a reasonable timeframe. We leverage DAG caching and breaking up the circuit into subcircuits to reduce the compilation time to only a couple of minutes, much less time than it would take to run on the physical hardware we consider in this paper. This indicates our block scheduler could be scaled for use on near-term hardware, especially if techniques like semi-classical QFT are used to decrease the compilation cost of QFT~\cite{griffiths1996semiclassical}. Together, these benchmarks show that the block scheduler produces high-quality schedules quickly enough to support routine resource-estimation studies across a wide range of computer sizes.


\subsection{Runtime scaling with computer size}
\noindent Lastly, in~\cref{fig:rt_against_logical_qubits} we simulate Qubitized Fermi--Hubbard (FH)~\cite{babbush2018encoding} by parts, on computers of various sizes by steadily increasing the number of interleaving modules, $n_{IM}$. The code distance determines the final number of logical qubits in the computer, which we then plot on the x-axis. For each sample point we attempt to simulate $2\times2$, $4\times4$, $6\times6$, $8\times8$, and $10\times10$ lattice sizes. Each simulation can fail by either running out of workspace (QOOM error) or by taking too long to complete at the selected code distance, such that the probability of a hardware fault compromises the computation, see~\cref{sec:resource_estimation}. We start by using the ARE model to determine a single-shot code distance, $d_{\mathrm{ARE}}$. Next, we decrement this code distance by 2 and try to simulate each lattice size from smallest to largest. If the scheduler fails, we increment the code distance and try again, stopping at $d_{\mathrm{ARE}}+2$. For the FH circuit, which is memory-heavy, we expect the scheduler to succeed by $d_{\mathrm{ARE}}$, since its reduced average memory occupancy provides substantially more workspace. Although ARE underestimates bridge and stale-state overheads, which partially offsets this advantage, this effect is small compared with the memory reduction observed here. The extra two-distance margin is therefore used only as a conservative safeguard.

We see diminishing returns in runtime reductions for all lattice sizes as the computer grows in size. Each lattice size has the same memory requirements for data qubits, so any additional logical qubits increase the workspace size. However, each additional qubit represents a smaller fraction of the current workspace size and becomes less impactful. Therefore, the diminishing returns effect is expected. Interestingly, the $8\times8$ lattice runtimes are faster than $6\times6$, although $8\times8$ requires a larger initial computer size. This is because of AV optimizations available for lattice sizes that are powers of 2, which means $8\times8$ has a lower AV than $6\times6$, at $3.28 \times 10^5$ and $3.36 \times 10^5$ respectively. $8\times8$ requires a larger minimum computer size because it has a larger QPE register, so its data qubit memory requirements are higher, which leaves less workspace on a computer of a given size. As a result, increasing the number of logical qubits yields greater runtime improvement for $8\times8$ than for $6\times6$, i.e., the diminishing returns effect explained earlier.


\FloatBarrier
\section{Conclusion}
\label{sec:conclusion}

\noindent We presented the Block Scheduler, a tool for Active Volume (AV) quantum architectures that schedules gates in time and explicitly allocates qubits as data qubits, workspace, bridge qubits, stale state storage, or unused qubits at the level of logical cycles. This cycle-resolved model allows us to test assumptions underlying prior analytic resource estimation (ARE) methods, which treat bridge and stale state (BSS) qubits through a fixed overhead and assume reaction latency is small such that stale states do not accumulate across logical cycles. By modeling qubit usage and stale state lifetimes explicitly, the scheduler reveals how memory, BSS overhead, reaction limits, and available workspace depend on computer size.

The main result of this work is that explicit scheduling reveals a workspace-dependent transition to reaction-limited execution in AV architectures. In the reduced-precision full-circuit $4\times4$ Fermi--Hubbard test circuit studied here, this transition occurs at approximately $10^3$ logical qubits. Here it is common for stale states to outlive a single logical cycle, and the average workspace subsequently changes from strong linear growth with computer size to a much shallower trend. This shows that reaction latency and workspace are coupled. Adding logical qubits initially exposes more parallelism and reduces runtime, but this same parallelism produces stale states more quickly. Once stale states cannot be cleared within a logical cycle, they persist in memory and restrict the effective workspace available in later cycles.

This transition has important consequences for AV resource estimation. Analytic estimates that neglect reaction latency are valid only below the reaction-limited regime. Above this regime, reaction latency is not a small fixed overhead: it changes the effective workspace available to the computation and therefore changes how runtime scales with computer size. This is the key execution effect captured by the scheduler but missed by previous ARE models.

Below the reaction-limited regime, the scheduler provides a direct test of the assumptions used in ARE. For the full-precision $4\times4$ Fermi--Hubbard instance considered here, the scheduler predicts a $1.64\times$ shorter runtime than ARE. The dominant source of this difference is the memory model: ARE uses the qubit high-water mark, while the scheduler tracks the data qubit requirement cycle by cycle and finds a substantially smaller average memory occupancy. The resulting increase in available workspace reduces the number of logical cycles and, in this instance, is large enough to lower the required code distance. Specifically, the average memory occupancy falls from $51.9\%$ to $28.8\%$ of the logical qubits, and the code distance drops from $33$ to $32$.

The scheduler also shows that bridge and stale state overhead should not be modeled as a fixed fraction of memory. For a fixed circuit, the ARE model predicts that this overhead fraction decreases with computer size, whereas the scheduler shows the opposite trend: larger workspaces allow more operations to be executed per logical cycle, increasing both bridge demand and stale state production. The fitted large-workspace asymptote is numerically close to the $20\%$ factor used in~\cite{Litinski22Active}, but this agreement should be viewed as coincidental. In the scheduler, the asymptote reflects a maximum-bridging regime, not a memory-proportional overhead.

The block scheduler exposes routing constraints that are hidden by aggregate resource estimates. For the test circuit, routing data qubits that participate in different operations in consecutive logical cycles can be accommodated using an average of only 1.36 additional bridge qubits per cycle, suggesting negligible overhead in the EFTQC regime. Workspace routing may become more important as the workspace fraction increases. In our conservative model, up to approximately 30\% of the computer must be rearranged in the average case and 50\% in the worst case between logical cycles. Although these estimates assume that all required rearrangement occurs between cycles and therefore provide pessimistic upper bounds, they indicate that workspace placement and reconfiguration may become an important constraint in large, workspace-dominated AV computers.

The scheduler also appears practical for resource-estimation studies. The MILP comparison gives an exact optimality certificate on a small benchmark and evidence of near-optimality on a larger instance, while component-wise scheduling allows larger circuits to be studied when full-DAG construction is impractical. These properties make the block scheduler a natural foundation for the full compilation pipeline envisioned for Active Volume architectures.

Several directions remain for future work. First, the reaction-dominant transition and the workspace-dependent BSS model should be tested across a broader benchmark suite and on more optimized circuit constructions. Second, the scheduler should be extended from a time-scheduling tool to a joint time-and-space scheduler that explicitly models qubit placement and routing, particularly in the large-workspace regime identified here. This would allow hardware quantities such as the required routing capability and fusion distance $r$ to be extracted directly from scheduled computations and enable comparisons between alternative routing architectures and spacetime tradeoffs. Finally, optimized high-level subroutine libraries could be integrated into the scheduler, allowing common operation patterns to be replaced by lower-AV composite blocks and potentially reducing bridge overhead. Together, these extensions would result in a full-stack qubit-placement and routing tool for AV hardware, enabling end-to-end resource estimates that remain both accurate and computationally tractable. Such a tool would give algorithm designers and hardware architects a shared, cycle-accurate picture of how their choices shape the performance of Active Volume computers.

\begin{acknowledgments}
\noindent Thanks to Dylan Sim, William Simon, and Kevin Gui for their help with circuit implementation and preparation. Thanks to Sam Pallister, William Pol, and Daniel Litinski for helpful comments.
\end{acknowledgments}

\section*{Author Contributions}
Sam Heavey contributed the theory of the block scheduler model, the development of the scheduler code, the MILP solver, the $\ket{Y}$ state handling methods and their analysis, the extensions to the DAG construction code that support the block scheduler, and the validation of the scheduling-by-parts method. Athena Caesura contributed the core development of the DAG construction code, the qubit routing theory and analysis, the circuit compilation, and the speedups to the block scheduler.

The authors used AI to assist with rewording and reformatting the paper, plotting results, and creating the interface to the ILP solver.

\bibliographystyle{quantum}
\bibliography{references}

\appendix

\section{Magic State Injection}
\label{sec:magic_states}

\noindent The active volume formalism typically subsumes the cost of magic-state injection into the cost of the corresponding gate itself. Although this is useful for obtaining rapid resource estimates, it can obscure important details. In this section, we examine the magic-state injection process more closely in order to identify precisely where the associated costs arise and how the formalism behaves when gates are bridged.

\subsection{Careful cost accounting of magic state injection}

\begin{figure}
    \centering
    \includegraphics[width=\linewidth]{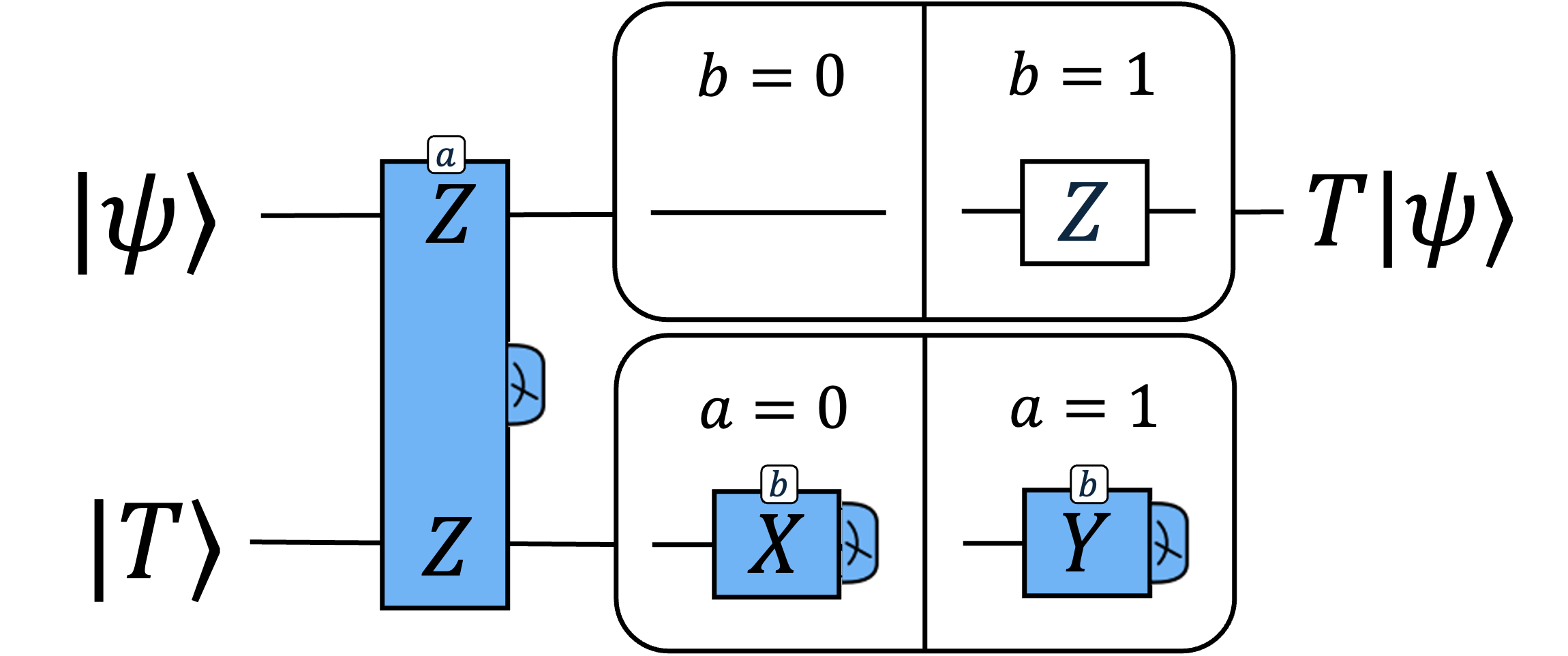}
    \caption{The magic state injection scheme used in this paper. A ZZ measurement is performed, producing an outcome $a$. Once this ZZ measurement is performed we now refer to the magic state as a stale state. $a$ is then used to determine the basis in which the magic state is measured. The outcome of the measurement of the magic state $b$ is then used to determine whether or not a Z correction is applied.}
    \label{fig:magic_state_injection}
\end{figure}

\noindent In this paper, we use stale states, sometimes also called auto-corrected states~\cite{fowler2012time, litinski2019game}, rather than traditional phase-gate corrections. This concept arises naturally from the injection scheme shown in~\cref{fig:magic_state_injection}, which allows computation to continue on the qubit into which the state was injected. Within the active volume architecture, this is particularly advantageous, as it permits continuous computation on that qubit without waiting for the correction to be applied.

\begin{figure}
    \centering
    \includegraphics[width=\linewidth]{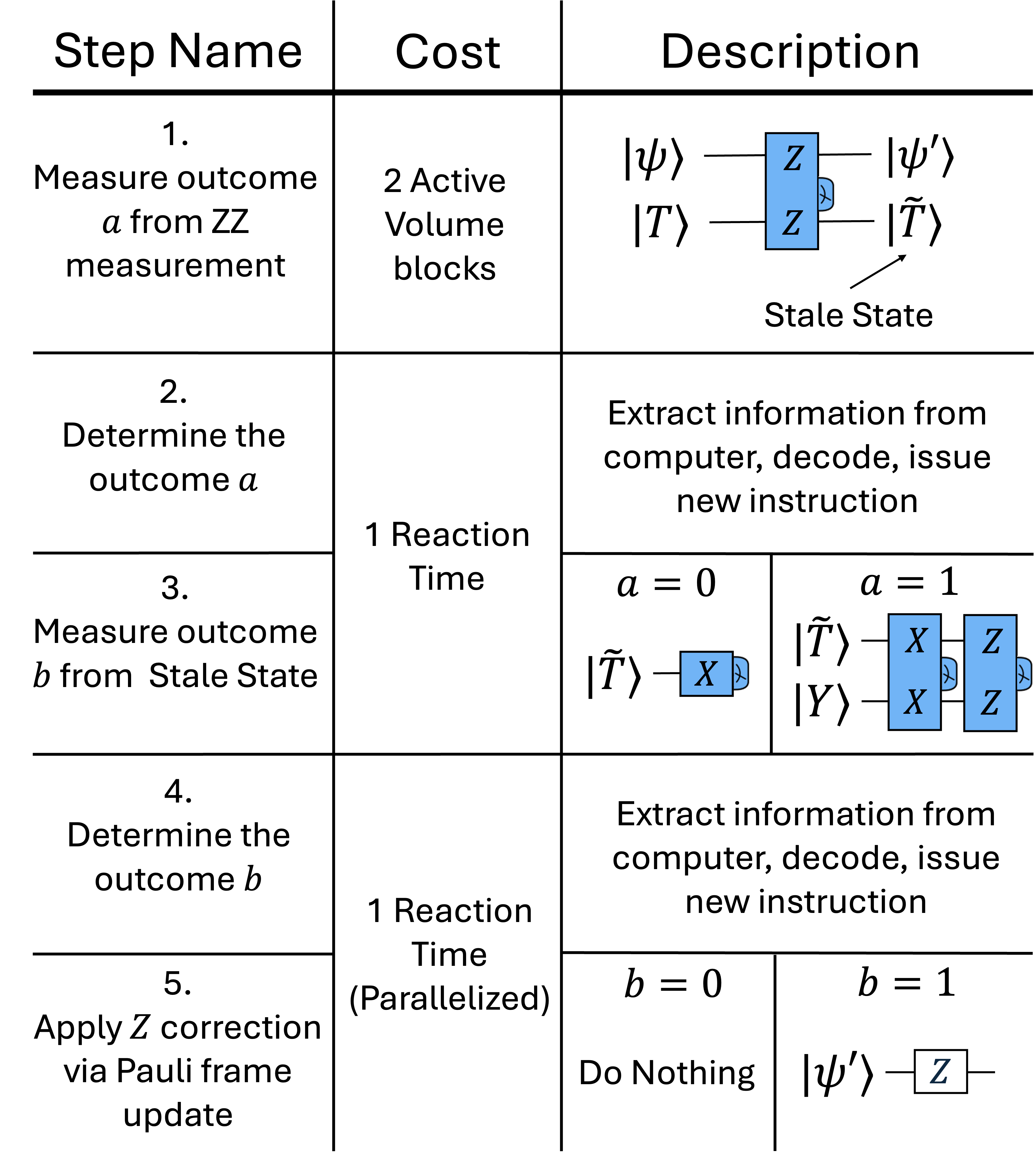}
    \caption{A table showing a detailed analysis of each step in the magic-state injection protocol used in this paper. Although the final two steps may appear to introduce additional latency, they can be executed in parallel with other reactions as they are independent, yielding a total cost of two Active Volume blocks and one reaction time of idling for the stale state.}
    \label{fig:magic_state_injection_cost}
\end{figure}

To account for the time cost of this circuit, we proceed step by step in~\cref{fig:magic_state_injection_cost}. First, we perform a $ZZ$ measurement between the qubit and the magic state. This step is straightforward to account for, as it corresponds to the execution of Active Volume. After this step, the magic state is referred to as a \emph{stale state}, since it is  waiting to be measured. Specifically, it remains idle until Step~2 and 3 have taken place.

In Step~3, we implement the measurement basis chosen in Step~2, either by a direct \(X\)-basis measurement or by a transversal Bell measurement between the stale state and a \(\ket{Y}\) state. Either operation can begin as soon as the basis is known and completes within one code cycle, even when initiated partway through a logical cycle. 

We define the reaction time as the total time required to perform Steps 2 and 3, that is the time take to: determine the measurement outcome $a$, communicate the next measurement instructions, and perform this measurement. We assume a reaction time of \(15\,\mu\mathrm{s}\), including an approximately \(10\,\mu\mathrm{s}\) measurement time obtained for \(d=30\) using the hardware parameters in~\cref{tab:hardware_parameters}. This is a conservative assumption, in line with the estimates in~\cite{Litinski22Active}.

If one wished to reduce the stale state measurement time below one code cycle, an additional switch could route stale states into a shorter delay line, potentially reducing the measurement time to the nanosecond scale~\cite{Litinski22Active}. For \(d=30\), Step~3 is expected to dominate the runtime, and could in principle be executed about \(25\) times before reaction limited behavior begins.

At first glance, it may seem that Steps~4 and~5 require an additional reaction time. It is reasonable to exclude Step~5 from the runtime, since it is merely a classical update that can be tracked during computation. But without further explanation it may not be immediately obvious why Step~4 does not incur an additional reaction time cost. This will become clearer in the next section where we consider the case of bridging two non-commuting non-Clifford gates and show it can just be pushed into the next reaction layer.

\subsection{Parallelizing the Active Volume of non-Clifford gates}

\noindent When parallelizing across non-Clifford gates, it is important to distinguish between the AV portion of the computation and the associated reactions. Only the AV can be parallelized; the reactions themselves cannot be parallelized and must instead be executed sequentially. Fortunately, as explained in the previous section, for the architecture considered in this paper these reactions are approximately $25$ times faster than the execution of the active volume.

\begin{figure*}[tp]
    \centering
    \includegraphics[width=.69\linewidth]{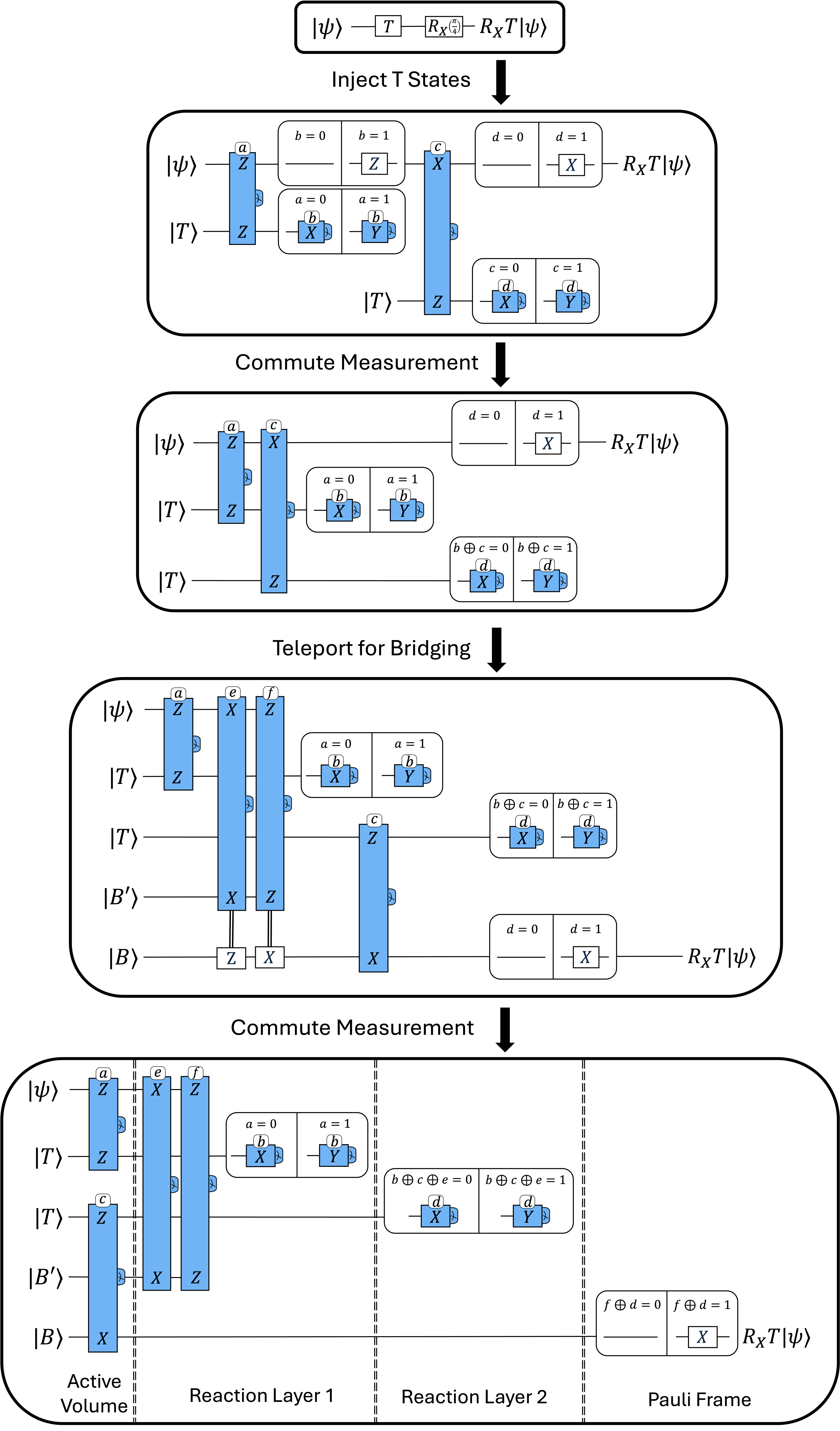}
    \caption{A series of compilation steps showing how exactly one would execute the active volume of 2 non-commuting, non-Clifford gates in the same logical cycle in the Active Volume architecture. In these circuit diagrams, results of each measurement are stored in a variable given at the top of the measurement and conditional gates are shown in boxes. In the bottom diagram, double dashed lines represent when reactions must occur in the circuit. In the first ``Commute Measurement" step, note that we included the outcome of $b$ in the box that was dependent on $c$ before. Recall from~\cref{fig:magic_state_injection_cost} that the $Y$ measurements can be performed in a single syndrome check using a Bell measurement. The bottom diagram shows the resulting circuit separated into active volume, reaction layers, and a Pauli frame.}
    \label{fig:multiple_injections}
\end{figure*}

In~\cref{fig:multiple_injections}, we provide a more precise description of how the active volume can be parallelized across multiple non-commuting non-Clifford gates. As shown in the figure, only three reactions are required in the circuit, and the final reaction can be parallelized with the active volume executed in the subsequent logical cycle. Therefore, within the logical cycle under consideration, only two reactions must be completed.

It is also notable that the Bell measurement can be parallelized with the conditioned $Y$ measurement. This implies that the bridging operation does not increase the number of reaction layers. If this construction is repeated for additional non-commuting non-Clifford gates, the same pattern persists. The situation for $\ket{CCZ}$ states is largely the same, except that the correction becomes a reactive $CZ$ gate rather than a $Y$ measurement.

\section{\texorpdfstring{Cost derivations for simulating reactive $Y$ measurements}{Cost derivations for simulating reactive Y measurements}}
\label{sec:cost_derivation_reactive_y_methods}

\noindent In practice, reactive $Y$ measurements are used to implement Clifford corrections for $\ket{T}$ state injections \cite{Litinski22Active}. Therefore, they are associated with stale $\ket{T}$ states. The measurement outcomes related to $\ket{T}$ state injection start decoding as soon as a logical cycle is complete. During this time the $\ket{T}$ sits as a stale state. After we have finished decoding we learn whether a reactive $Y$ measurement is required or if we can safely remove the stale states by measuring them in the $X$ basis, where either outcome has a 50\% probability. In this section we talk exclusively about $\ket{Y}$ state production and reactive $Y$ measurements, however, the analysis extends to any reactive measurement that requires magic states for corrections (e.g. reactive $T$).

Ideally, we would produce the exact number of $\ket{Y}$ states required by reactive measurements in the next logical cycle in the current logical cycle. However, this number is not known when the production decision is made. The next logical cycle begins before the relevant measurement outcomes have been decoded, and stale state chains are resolved sequentially: the measurement basis for a stale state may only become known after the outcomes of earlier non-commuting measurements, including earlier reactive $Y$ corrections, are decoded.

We therefore choose a production target for each logical cycle. If too many $\ket{Y}$ states are produced, the surplus states become spares and must be stored until they are consumed by later gates. If too few are produced, then at least one stale state whose correction requires a $\ket{Y}$ state must remain alive until a new $\ket{Y}$ state is produced. Over-production therefore costs idle $\ket{Y}$ state spacetime volume, while under-production add storage costs due to delayed stale state removal. The optimal production target balances these two costs.

Let $\mathcal{S}_\ell$ be the set of stale states who may require a reactive $Y$ measurement in logical cycle $\ell+1$, and let $k_\ell=|\mathcal{S}_\ell|$. We index the production decision by the logical cycle $\ell$ in which the $\ket{Y}$ states are prepared. For each stale state $i\in\mathcal{S}_\ell$, define the Bernoulli random variable
\[
B_{\ell,i} =
\begin{cases}
1, & \text{if } i \text{ requires a reactive}\\
 & \text{$Y$ measurement},\\
0, & \text{otherwise},
\end{cases}
\]
with $\Pr(B_{\ell,i}=1)=p_{\ell,i}$. The total $\ket{Y}$ state demand generated by these stale states is therefore
\begin{equation}
X_\ell=\sum_{i\in\mathcal{S}_\ell} B_{\ell,i}.
\label{eq:reactive_y_demand}
\end{equation}
In general, $X_\ell$ follows a Poisson-binomial distribution,
\begin{equation}
\Pr(X_\ell=x) = \sum_{\substack{A\subseteq \mathcal{S}_\ell\\ |A|=x}} \prod_{i\in A}p_{\ell,i} \prod_{j\in \mathcal{S}_\ell\setminus A}(1-p_{\ell,j}).
\label{eq:poisson_binomial_y_demand}
\end{equation}
Its mean and variance are
\begin{equation}
\begin{gathered}
\mu_\ell=\mathbb{E}[X_\ell]=\sum_{i\in\mathcal{S}_\ell}p_{\ell,i},\\
\sigma_\ell^2=\mathrm{Var}(X_\ell)=\sum_{i\in\mathcal{S}_\ell}p_{\ell,i}(1-p_{\ell,i}).
\end{gathered}
\label{eq:y_demand_mean_variance}
\end{equation}
For the case in which all relevant stale states arise from $\ket{T}$ state injections, $p_{\ell,i}=1/2$ and $X_\ell\sim\mathrm{Binomial}(k_\ell,1/2)$.

We parameterize the number of $\ket{Y}$ states produced in logical cycle $\ell$ as
\begin{equation}
N_\ell(\eta) = \min\!\left\{ k_\ell,\, \max\!\left\{ 0,\, \left\lceil \mu_\ell+\eta\sigma_\ell\right\rceil \right\} \right\},
\label{eq:eta_policy}
\end{equation}
where $\eta$ is a global tuning parameter. The value $\eta=0$ corresponds to producing approximately the expected demand, $\eta>0$ biases the scheduler toward over-production, and $\eta<0$ biases it toward under-production.

For a fixed production target $N_\ell$, define the surplus and deficit random variables
\begin{equation}
O_\ell(N_\ell)=(N_\ell-X_\ell)^+, \quad U_\ell(N_\ell)=(X_\ell-N_\ell)^+,
\label{eq:surplus_deficit}
\end{equation}
where $(a)^+=\max(a,0)$. Their expectation values are
\begin{equation}
\mathbb{E}[O_\ell(N_\ell)] = \sum_{x=0}^{N_\ell}(N_\ell-x)\Pr(X_\ell=x),
\label{eq:expected_surplus}
\end{equation}
and
\begin{equation}
\mathbb{E}[U_\ell(N_\ell)] = \sum_{x=N_\ell+1}^{k_\ell}(x-N_\ell)\Pr(X_\ell=x).
\label{eq:expected_deficit}
\end{equation}
The expected first-order cost of the policy is then
\begin{equation}
\begin{aligned}
C(\eta;n_q) = \sum_\ell \Bigl[&N_\ell(\eta) + \tau_I(n_q)\mathbb{E}\!\left[O_\ell(N_\ell(\eta))\right] \\
&+ \tau_D(n_q)\mathbb{E}\!\left[U_\ell(N_\ell(\eta))\right]
\Bigr].
\end{aligned}
\label{eq:reactive_y_general_cost}
\end{equation}
Here $n_q$ is the number of logical qubits assigned to the computation, $\tau_I(n_q)$ is the average lifetime of a surplus $\ket{Y}$ state measured in logical cycles, and $\tau_D(n_q)$ is the average aggregate delay cost caused by one missing $\ket{Y}$ state. The latter includes the extra time for which the directly affected stale state remains live, as well as any additional stale state lifetime induced downstream in a reaction chain.

For intuition, when $k_\ell$ is large we may approximate $X_\ell$ by a normal distribution, $X_\ell\approx\mathcal{N}(\mu_\ell,\sigma_\ell^2)$. Ignoring integer rounding and writing
\[
a_\ell=\frac{N_\ell-\mu_\ell}{\sigma_\ell},
\]
we obtain
\begin{equation}
\begin{gathered}
    \mathbb{E}[O_\ell] \approx \sigma_\ell\left[a_\ell\Phi(a_\ell)+\phi(a_\ell)\right],\\
    \mathbb{E}[U_\ell] \approx \sigma_\ell\left[\phi(a_\ell)-a_\ell(1-\Phi(a_\ell))\right],
\end{gathered}
\label{eq:normal_surplus_deficit}
\end{equation}
where $\Phi$ and $\phi$ are the standard normal CDF and PDF. In particular, when $N_\ell=\mu_\ell$,
\begin{equation}
\mathbb{E}[O_\ell]\approx \mathbb{E}[U_\ell]\approx \frac{\sigma_\ell}{\sqrt{2\pi}}.
\label{eq:half_normal_surplus_deficit}
\end{equation}
For $p_{\ell,i}=1/2$, this gives
\begin{equation}
C(0;n_q)
\approx
\sum_\ell
\left[
\frac{k_\ell}{2}
+
\left(\tau_I(n_q)+\tau_D(n_q)\right)
\sqrt{\frac{k_\ell}{8\pi}}
\right],
\label{eq:expected_y_policy_cost}
\end{equation}
or equivalently, if $N_\ell=k_\ell/2$ denotes the expected number of produced $\ket{Y}$ states,
\[
C(0;n_q)
\approx
\sum_\ell
\left[
N_\ell
+
\left(\tau_I(n_q)+\tau_D(n_q)\right)
\sqrt{\frac{N_\ell}{4\pi}}
\right].
\]

Since $\tau_I$ and $\tau_D$ depend on the circuit structure, the qubit allocation, and the resulting stale state geometry, we determine the best value of $\eta$ by simulation. For each logical qubit count $n_q$, we evaluate the runtime over a range of $\eta$ values and choose the minimum. Note we use the runtime rather than evaluating \cref{eq:reactive_y_general_cost}, since this avoids extracting $\tau_I$ and $\tau_D$, and minimizing runtime is the objective of our optimization.

\begin{figure*}
    \centering
    \includegraphics[width=.83\linewidth]{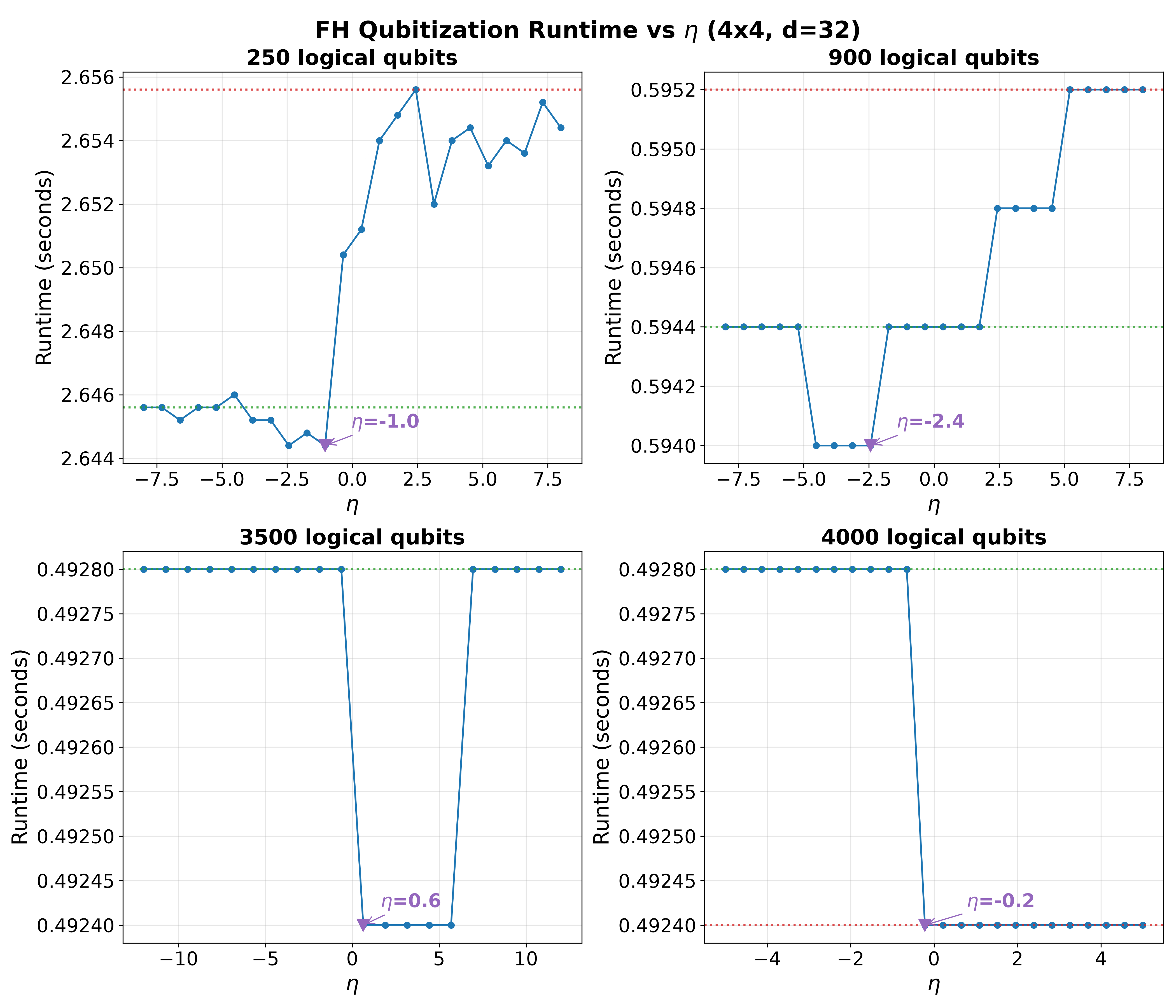}
    \caption{The total runtime as a function of the production-bias parameter $\eta$ for four representative logical qubit counts. We schedule the full DAG of the $4\times4$ qubitized Fermi--Hubbard circuit (\cref{sec:test_circuit_description}) with a half-precision phase register, since realistic runtimes are not required for this analysis and the reduced precision keeps DAG generation tractable. (see \cref{subsec:reaction_dominant_regime}). The $\ket{Y}$ state production target for logical cycle $\ell$ is $N_\ell(\eta)=\lceil\mu_\ell+\eta\sigma_\ell\rceil$, constrained to the interval $[0,k_\ell]$. Negative $\eta$ under-produces relative to the expected demand, whereas positive $\eta$ over-produces. The minimum runtime for each curve is annotated in purple; when multiple minima exist, we select the $\eta$ value closest to zero. The dotted red and green lines denote the two limiting strategies: producing the maximum possible number of $\ket{Y}$ states for reactive measurements and producing none, respectively. At low logical qubit counts, asymptotic over-production performs worse than under-production, but this trend reverses as the computer size increases, with the crossover occurring at approximately 3500 logical qubits. The overall variation in runtime is minimal especially at larger qubit counts. Therefore, for simplicity one may consider always doing maximum production which performs better at low qubit counts.}
    \label{fig:reactive_y_eta_cost_curves}
\end{figure*}

The change in the optimal value of $\eta$ reflects a transition between two regimes. When stale state chains are shallow, the cost of a missing $\ket{Y}$ state is often local: the affected stale state waits, but few or no later states are delayed. In this regime, it can be beneficial to produce slightly fewer than the expected demand, because unused $\ket{Y}$ states may idle for several logical cycles. As the available logical qubit count increases, the scheduler can support longer stale state chains. The computation then becomes increasingly reaction dominated, and a single missing $\ket{Y}$ state can delay not only the stale state that requires the correction, but also later stale states whose bases depend on that correction outcome. This increases $\tau_D(n_q)$ and shifts the optimum toward producing more $\ket{Y}$ states in advance.

\subsection{Limiting production strategies}
\label{sec:reactive_y_limiting_strategies}

\cref{fig:reactive_y_eta_cost_curves} can be better understood by considering the two extremes of $\ket{Y}$ state production. The first extreme is maximum production: produce a $\ket{Y}$ state for every stale state that might require one. This corresponds to $N_\ell=k_\ell$. The cost is
\begin{equation}
C_1=\sum_\ell\left[k_\ell+\tau_I(n_q)\mathbb{E}[k_\ell-X_\ell]\right].
\label{eq:max_y_production_general}
\end{equation}
For $p_{\ell,i}=1/2$, $\mathbb{E}[k_\ell-X_\ell]=k_\ell/2$, so
\begin{equation}
\mathbb{E}[C_1] = \sum_\ell k_\ell \left(1+\frac{\tau_I(n_q)}{2}\right).
\label{eq:max_y_production_half_probability}
\end{equation}
Equivalently, if
\[
N_{\mathrm{poss}}=\sum_\ell k_\ell
\]
is the total number of possible reactive $Y$ measurements, then
\begin{equation}
\mathbb{E}[C_1] = \left(1+\frac{\tau_I(n_q)}{2}\right)N_{\mathrm{poss}}.
\label{eq:method_1_limit}
\end{equation}
This strategy avoids reactive $Y$ measurement delays, but it dedicates workspace to $\ket{Y}$ states that may not be needed in the next logical cycle. On average, half of these states become spares in the $p=1/2$ case.

The opposite extreme is minimum production: produce no $\ket{Y}$ states in advance, and only create a $\ket{Y}$ state after the decoder determines that one is required. This corresponds to $N_\ell=0$. The cost is then dominated by stale state delay rather than idle $\ket{Y}$ states. For a simple stale state chain in which each stale state has one predecessor, the expected additional lifetime of a stale state with reaction layer $R$ is
\begin{equation}
f(R) = \frac{1}{2}\left[2(R-1)+1\right] = R-\frac{1}{2},
\label{eq:reaction_layer_delay_cost}
\end{equation}
for the $p=1/2$ case. The term $1/2$ is the probability that the stale state itself requires a delayed $\ket{Y}$ correction, which adds one logical cycle. Each earlier stale state in the chain requires a delayed correction with probability $1/2$, and each such event adds two logical cycles to the lifetime of the downstream stale state: one cycle to create and apply the $Y$ measurement, and one further cycle before the correction outcome has been decoded and propagated.

The minimum-production delay cost is therefore
\begin{equation}
C_2= \sum_i f(R_i),
\label{eq:method_2_limit_sum}
\end{equation}
or, using an average reaction layer $\bar{R}$,
\begin{equation}
\mathbb{E}[C_2] \approx N_{\mathrm{poss}} f(\bar{R}).
\label{eq:method_2_limit_average}
\end{equation}
This expression is optimistic for stale states with multiple predecessors, because any predecessor requiring a correction can force the downstream state to remain live.

In the $p=1/2$ case the cost of not having a $\ket{Y}$ state prepared for a stale state is $f(R=1)$. We can compare this to the cost of preparing a state in advance:
$$f(R=1)=\frac{1}{2}<1+\frac{\tau_I(n_q)}{2}.$$
So when stale states are terminal, i.e., delaying their correction does not delay other stale states it is always cheaper to make a $\ket{Y}$ state after decoding. This explains the location of the dip in the optimal $\eta$ curve. A slightly negative optimal $\eta$ means that the scheduler intentionally produces fewer $\ket{Y}$ states than the expected demand. This is likely to leave terminal stale states without a $\ket{Y}$ state but as these are cheaper to handle on demand, thus reducing the overall cost.

\section{Qubit Routing Models}
\label{sec:qubit_routing_models}

\subsection{Workspace layout and inter-cycle routing}

When the number of workspace qubits is less than or comparable to the number of data qubits, routing is relatively simple. For example, the two-row configuration of~\cite{Litinski22Active} can keep the workspace clustered while ensuring that it remains accessible to the data qubits. This arrangement becomes less practical once workspace qubits outnumber data qubits, since the workspace itself may need to be reconfigured to accommodate different sized operations.

Consider a workspace-dominated computer in which a logical cycle containing only Hadamard operations is followed by one containing only CCZ preparations. The first cycle may partition the workspace into many small regions, whereas the second requires fewer, much larger contiguous regions. Since workspace cannot be rearranged during a logical cycle, some workspace qubits may therefore need to be rearranged before the operations in the next cycle can begin. Moreover, routing data qubits may become difficult during logical cycles in which workspace occupies most of the machine. Because workspace qubits participating in lattice surgery cannot be moved during those cycles, they may substantially restrict the routes available to the data qubits.

For the routing-demand estimates below, we assume that all required rearrangement occurs between logical cycles, after workspace qubits have been released. This is the conservative inter-cycle model used in~\cref{subsubsec:qubit_routing}.

\subsection{Routing-demand models}

\noindent For each logical cycle, the block scheduler partitions the qubits into five classes: data qubits, workspace qubits, bridge qubits, stale states, and unused qubits. We first characterize the routing properties of each class. We then use the greedy scheduler to determine the number of qubits in each class required at every logical cycle.

Workspace qubits and bridge qubits are drawn from a common fungible pool of unused qubits at the beginnning of each logical cycle and are released at the end of the logical cycle. We therefore treat these two qubit types identically in the routing model. An operation requires suitable workspace at a particular location, but not a particular workspace qubit. Consequently, a scheduled interaction does not require workspace to be moved if suitable workspace is already present at the required location. Data qubits are non-fungible, however only a subset of them must be routed in a given logical cycle, namely those which are acted upon by a gate. The greedy scheduler also stores the number of data qubits which are used by operations. Stale states and unused qubits never need to be routed, but they still occupy space in the machine, much like data qubits that do not need to be moved in a given cycle.

Let \(d_\ell\) denote the number of data qubits that must be routed in logical cycle \(\ell\), and let \(w_\ell\), \(b_\ell\), \(s_\ell\), and \(u_\ell\) denote the numbers of workspace qubits, bridge qubits, stale states, and unused qubits, respectively, in that cycle. Let \(N\) denote the total number of qubits in the machine. It is convenient to define the number of active fungible roles in cycle \(\ell\) as
\begin{equation}
a_\ell = w_\ell + b_\ell,
\end{equation}
and the size of the fungible pool as
\begin{equation}
p_\ell = w_\ell + b_\ell + u_\ell = a_\ell + u_\ell.
\end{equation}

\begin{figure}
    \centering
    \includegraphics[width=\linewidth]{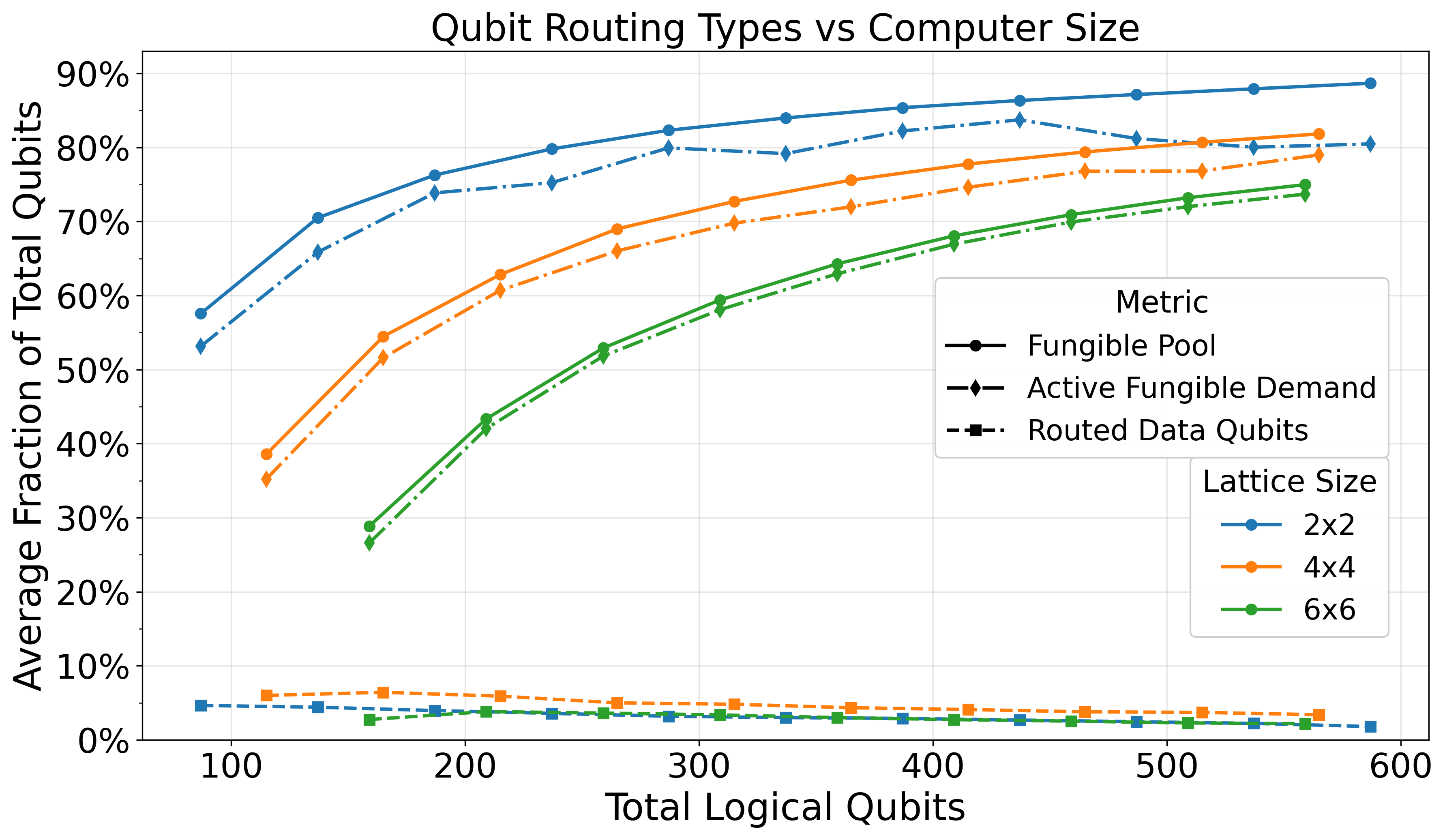}
    \caption{
    Average active fungible demand, fungible pool size, and routed data-qubit fraction as a function of machine size for several lattice sizes in the qubitized Fermi--Hubbard simulation. Specifically, we plot the averages of \((w_\ell+b_\ell)/N\), \((w_\ell+b_\ell+u_\ell)/N\), and \(d_\ell/N\) over all logical cycles, where $N$ is the total number of logical qubits. The gap between the active fungible demand and the fungible pool illustrates the flexibility provided by fungibility, while the routed data-qubit fraction remains comparatively small across the machine sizes considered.
    }
    \label{fig:fh_routing_fractions_vs_computer_size}
\end{figure}

The routed fraction in cycle \(\ell\) is modeled as the sum of two contributions: the \(d_\ell\) data qubits that must be moved to realize the scheduled interactions, and the number of fungible qubits that must be routed to populate the \(a_\ell\) required active roles. In~\cref{fig:fh_routing_fractions_vs_computer_size}, we plot the average values on an \(N\)-qubit machine of \(a_\ell/N\), \(p_\ell/N\), and \(d_\ell/N\) over all logical cycles for several lattice and machine sizes. As expected, the active fungible demand remains below the total fungible pool, while the fraction of routed data qubits remains comparatively small. This size gap is what allows fungibility to reduce the routing demand in both the worst-case and average-case models below.

\paragraph{Worst-case model.}
In the worst-case model, we conservatively assume that all \(d_\ell\) participating data qubits must be routed. For the fungible qubits, the worst case is obtained by minimizing the overlap between the \(a_\ell\) required active locations and the fungible pool of size \(p_\ell\). Any two subsets with sizes \(a_\ell\) and \(p_\ell\) must overlap on at least
\begin{equation}
\max\!\bigl(0,\, a_\ell + p_\ell - N \bigr)
\end{equation}
qubits. Therefore, the maximum number of fungible qubits that must be routed in cycle \(\ell\) is
\begin{equation}
\begin{aligned}
r_{\ell,\mathrm{f}}^{\mathrm{worst}}
&=a_\ell - \max\!\bigl(0,\, a_\ell + p_\ell - N \bigr)\\
&=\min(a_\ell,\, N-p_\ell).
\end{aligned}
\end{equation}
The resulting worst-case routed fraction in cycle \(\ell\) is
\begin{equation}
f_\ell^{\mathrm{worst}}
=
\frac{d_\ell + \min(a_\ell,\, N-p_\ell)}{N}.
\end{equation}
Equivalently,
\begin{equation}
f_\ell^{\mathrm{worst}}
=
\frac{d_\ell + \min\!\bigl(w_\ell+b_\ell,\, N-w_\ell-b_\ell-u_\ell\bigr)}{N}.
\end{equation}
This expression makes the overlap constraint explicit. When the fungible pool occupies more than half of the machine, the required active locations must already overlap it non-trivially. As a result, not all workspace and bridge qubits can simultaneously be out of place, and the routing cost associated with the fungible qubits is bounded by the smaller non-fungible portion of the machine.

To summarize worst-case routing demand over an entire computation with \(n_{lc}\) logical cycles, we consider the maximum, over all logical cycles, of the routed fraction predicted by the worst-case model,
\begin{equation}
f_{\max}^{\mathrm{worst}} = \max_{\ell \in \{1,\dots,n_{lc}\}} f_\ell^{\mathrm{worst}}.
\end{equation}
This quantity identifies the single most routing-intensive cycle in the computation and is the quantity reported in~\cref{fig:qubit_routing_fraction} for the worst-case model.

\paragraph{Average-case model.}
In the average-case model, we again take all \(d_\ell\) participating data qubits to require routing, but now assume that the fungible pool is distributed uniformly throughout the machine before the start of cycle \(\ell\). Under this assumption, the expected overlap between the \(a_\ell\) required active locations and the fungible pool of size \(p_\ell\) is
\begin{equation}
\mathbb{E}[\mathrm{overlap}]
=
a_\ell \frac{p_\ell}{N}.
\end{equation}
Therefore, the expected number of fungible qubits that must be routed in cycle \(\ell\) is
\begin{equation}
r_{\ell,\mathrm{f}}^{\mathrm{avg}}
=
a_\ell - a_\ell\frac{p_\ell}{N}
=
a_\ell\left(1-\frac{p_\ell}{N}\right).
\end{equation}
The corresponding average-case routed fraction in cycle \(\ell\) is
\begin{equation}
f_\ell^{\mathrm{avg}}
=
\frac{d_\ell + a_\ell\left(1-\frac{p_\ell}{N}\right)}{N}.
\end{equation}
Equivalently,
\begin{equation}
f_\ell^{\mathrm{avg}}
=
\frac{d_\ell + (w_\ell+b_\ell)\left(1-\frac{w_\ell+b_\ell+u_\ell}{N}\right)}{N}.
\end{equation}

To summarize average-case routing demand over an entire computation with \(n_{lc}\) logical cycles, we consider the mean and standard deviation of the routed fraction predicted by the average-case model,
\begin{equation}
\bar{f}^{\mathrm{avg}} = \frac{1}{n_{lc}}\sum_{\ell=1}^{n_{lc}} f_\ell^{\mathrm{avg}},
\end{equation}
and
\begin{equation}
\sigma_f^{\mathrm{avg}}
=
\sqrt{
\frac{1}{n_{lc}}\sum_{\ell=1}^{n_{lc}}
\left(f_\ell^{\mathrm{avg}} - \bar{f}^{\mathrm{avg}}\right)^2
}.
\end{equation}
These quantities characterize the typical routing demand and its cycle-to-cycle variation across the computation, and are the quantities reported in~\cref{fig:qubit_routing_fraction} for the average-case model. Together, the worst-case and average-case models provide a conservative estimate and a typical estimate of the routing demand for each schedule.

\subsection{Runtime implications}
\label{sec:routing_runtime_implications}

Assuming that all routing is inter-cycle and the routing estimates of Sec.~6 of~\cite{Litinski22Active}, rearranging a computer with $10^6$ qubits may cost as many as approximately 50 code cycles of routing in the worst case and 30 code cycles in the average case. If all of this routing were placed between logical cycles it could therefore produce a substantial runtime overhead in workspace-dominated machines. So even with these pessimistic upper bounds, routing about halves our computational speed.

An important direction for future work is therefore to develop layouts and routing algorithms for the regime in which workspace qubits begin to outnumber data qubits. One simple approach would be to reserve additional unused qubits to maintain routing channels similar to those of~\cite{Litinski22Active}, trading additional hardware for reduced routing time. Exploring these spacetime tradeoffs may substantially reduce the routing overhead at only modest additional qubit cost.

\section{MILP Solver Specification}
\label{sec:MILP_specs}
\noindent Here we give more details on the MILP solver and how we used it in our optimality arguments. MILP solvers consist of 3 major components, variables which determine the size of the solution space, constraints which limit the region of the solution space that is explored, and an objective. In our case the objective is to minimize the number of logical cycles that are required to complete the computation. The variables and constraints are less trivial and so we provide them in their entirety. Note, large solution spaces are expensive to explore and lead to impractical runtimes, so in this setup we only model non-reaction dominant EFTQC regimes where the it can be assumed that stale states never persist for more that one logical cycle. This is justified as the solver is only practical for small circuits/DAGs, i.e., that of a small QROM (see \cref{sec:scheduler_benchmarks}). We therefore use a small hardware budget 200 logical qubits to avoid the trivial single-logical-cycle regime, in which all feasible schedules are equally optimal in depth. By forcing the computation to span multiple logical cycles, the MILP has more freedom to move operations between cycles and therefore more opportunity to find a schedule that improves on the greedy solution. The peak number of stale state layers for computer with fewer than 200 logical qubits is too small to allow for stale states to last more than one logical cycle hence the validity of our simplification.

\begin{enumerate}
    \item Decision variables.
\end{enumerate}
\begin{itemize}
    \item $x_{v,\ell}\in\{0,1\}$. $x_{v,\ell}=1$ if vertex $v$ is scheduled into logical cycle $\ell$, otherwise $x_{v,\ell}=0$.
    \item $f_\ell\in\{0,1\}$. $f_\ell=1$ if logical cycle $\ell$ has a $\ket{Y}$ state factory, otherwise $f_\ell=0$.
    \item $y^{m}_{\ell}\in\mathbb{Z}_{\ge 0}$. $y^{m}_{\ell}$ represents the number of $\ket{Y}$ states created in logical cycle $\ell$.
    \item $y^{s}_{\ell}\in[0,1)$. $y^{s}_{\ell}$ denotes the residual fractional $\ket{Y}$ state inventory carried into logical cycle $\ell$.
    \item $s_\ell\in\mathbb{Z}_{\ge 0}$. $s_\ell$ represents the number of stale states occupying space in logical cycle $\ell$.
    \item $a_{q,\ell}\in\{0,1\}$. $a_{q,\ell}=1$ if qubit $q$ is still active at the end of logical cycle $\ell$, otherwise $a_{q,\ell}=0$
    \item $d_{q,\ell}\in\{0,1\}$. $d_{q,\ell}=1$ if qubit $q$ must be stored active at the end of logical cycle $\ell$, otherwise $d_{q,\ell}=0$
    \item $t_{q,\ell}\in\{0,1\}$. $t_{q,\ell}=1$ if qubit $q$ is used as an input to a gate scheduled into logical cycle $\ell$, otherwise $t_{q,\ell}=0$. 
    \item $m_{q,\ell}\in\{0,1\}$. $m_{q,\ell}=1$ if qubit $q$ is measured by a gate scheduled into logical cycle $\ell$, otherwise $m_{q,\ell}=0$.
\end{itemize}
\begin{enumerate}
    \item[2.] Constraints (high-level). The constraints enforce the variable definitions above:
\end{enumerate}
\begin{itemize}
    \item Each gate can only be scheduled into one logical cycle.
    \item Gates must be scheduled into the same logical cycle or a later logical cycle than their predecessors.
    \item If gates in the same logical cycle share qubits we must add a bridging qubit for each shared qubit.
    \item Qubits that have been activated in previous logical cycles, are not acted upon by gates in the current cycle, and have not been measured are allocated as data qubits.
    \item Stale states produced in the previous logical cycle must take up space in the current logical cycle.
    \item The sum of $\ket{Y}$ states made up to the current logical cycle must greater than or equal to the expected number of $\ket{Y}$ states consumed rounded up to the nearest integer.
    \item If a $\ket{Y}$ state is made in a logical cycle we must have a $Y$ factory in that logical cycle.
    \item If there is no $Y$ factory we allocate one qubit for the reserve catalsyt $\ket{Y}$ state.
    \item If the cumulative expected $\ket{Y}$ state consumption is less than the sum of $\ket{Y}$ states made, we allocate qubits for the expected $\ket{Y}$ state surplus.
    \item The total qubit usage each logical cycle must be less than the total number of qubits in the computer.
\end{itemize}

Together these variables and rules set a linear problem that can be brute force searched over all configurations by the MILP solver. To speed the search we use the greedy algorithms solution as a warm-start. To do this we converted the greedy algorithm's solution into a particular variable configuration and then had the solver proceed as normal.

\section{Test Circuit Description}
\label{sec:test_circuit_description}

\begin{figure*}[tp]
    \centering
    \includegraphics[width=\linewidth]{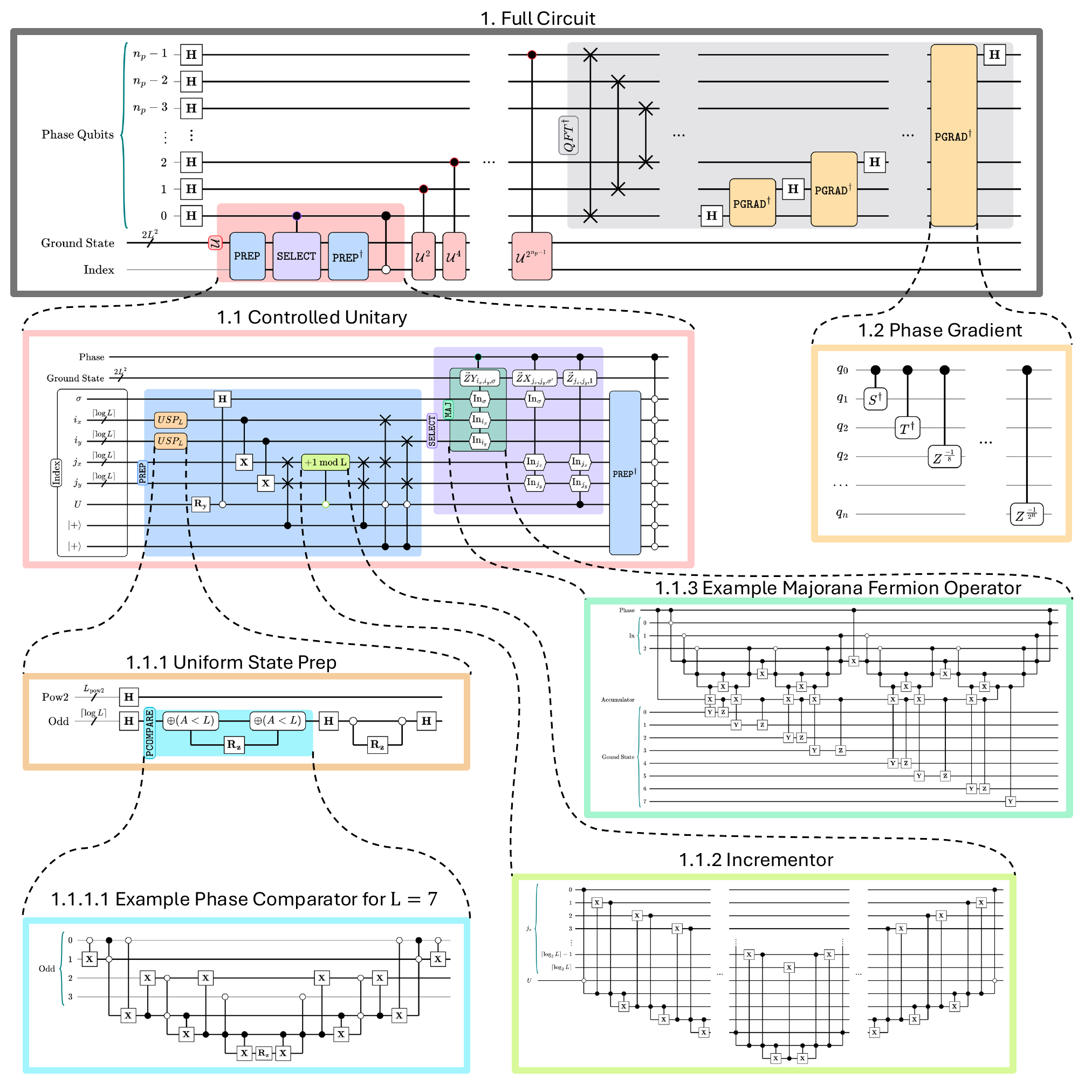}
    \caption{A circuit describing our implementation of QPE which is largely inspired by~\cite{kan2025resource} but simplifies the circuit at the cost of a slight increase in resource counts. It consists of 2 main subcircuits: controlled unitary $\mathcal{U}$ (red) and $QFT^\dagger$ is the inverse Quantum Fourier Transform (grey). Within each $\mathcal{U}$ there are smaller subcircuits: the uniform state prep on $L$ states USP$_L$, $+1 \text{mod L}$, and the Majorana Fermion operator. For the uniform state prep with $L$ states, one breaks up the circuit into $L=L_\text{pow2}L_\text{odd}$ where $L_\text{pow2}$ is a power of 2 and $L_\text{odd}$ is an odd number. $L_\text{pow2}$ and $L_\text{odd}$ define the sizes of the two registers of 1.1.1. Precise prescriptions for recreating the circuit are given in all panels except for 1.1.1.1 and 1.1.3 where examples are given instead. For precise prescriptions for 1.1.1.1 and 1.1.3, see~\cite{babbush2018encoding}. To implement~\texttt{PREP}$^\dagger$, note that 1.1.1.1 is an involution and the inverse of 1.1.2 is the $L - 1$ incrementor, which can be implemented in an analogous way that only differs from 1.1.2 by Paulis.}
    \label{fig:full_test_circuit}
\end{figure*}

\noindent We demonstrate our new resource estimation tools by compiling a Quantum Phase Estimation (QPE) circuit that estimates the ground state energy of the Fermi--Hubbard model using qubitization. Throughout this work, we set the Fermi--Hubbard coefficients to \(t=1\) and \(U=8\). Fermi--Hubbard simulation was chosen as an application because it is often cited as a target for computers in the early fault-tolerant quantum computing regime~\cite{kan2025resource, babbush2018encoding}. We use the QPE circuit given in~\cref{fig:full_test_circuit} as a testbed for our compilation algorithm which we call the \emph{test circuit}. We acknowledge that the test circuit is not state-of-the-art. Our goal here is not to report the latest or most optimized resource counts. Instead, we use a circuit whose gates are just representative of those expected in a benchmark application. This lets us isolate the effects of our compilation stack, quantify relative speedups, and test prior assumptions under conditions that are reasonably close to a useful application.

\subsection{Error budgeting}
\label{subsec:choice_of_parameters}

\noindent Our resource estimates distinguish between two sources of error: hardware error and algorithmic error. Hardware error arises from logical faults that survive quantum error correction, whereas algorithmic error arises from approximations made during circuit construction. We treat these sources separately. For every simulation, we choose the code distance so that the estimated probability of at least one hardware-induced logical fault during the computation is at most \(10\%\). We estimate the algorithmic error as the
sum of the errors caused by finite phase-register resolution and approximate rotation synthesis,
\begin{equation}
    \widehat{\epsilon}_{\mathrm{alg}}(L)
    =
    \epsilon_{\mathrm{phase}}(L)
    +
    \epsilon_{\mathrm{RS,tot}}(L),
    \label{eq:algorithmic_error_model}
\end{equation}
and choose the circuit parameters such that
\begin{equation}
    \widehat{\epsilon}_{\mathrm{alg}}(L) \leq 0.05.
    \label{eq:algorithmic_error_budget}
\end{equation}
This accounting neglects additional approximation errors in the construction of the controlled unitary and in converting the estimated phase to an energy. The lattice-dependent parameters used to control the algorithmic error are summarized in \cref{tab:circuit_parameters}.

\subsubsection{Phase-register truncation error}

\noindent
A finite phase register limits the precision with which quantum phase
estimation can resolve an eigenphase and, consequently, the corresponding
energy. We determine the number of phase qubits \(n_p\) using the expression
from~\cite{babbush2018encoding},
\begin{equation}
    n_p(L)
    =
    \left\lceil
    \log_2\left(
        \frac{\pi \lVert H(L)\rVert_1}
        {\sqrt{2}\epsilon_E(L)}
    \right)
    \right\rceil,
    \label{eq:number_of_phase_qubits}
\end{equation}
where \(\epsilon_E\) is the target precision in the ground-state energy and \(\lVert H\rVert_1\) is the 1-norm of the second-quantized Fermi--Hubbard
Hamiltonian.

To choose \(\epsilon_E\), we use the per-site energy accuracies reported for
classical methods in Table~II of~\cite{leblanc2015solutions}. For \(U=8\),
these accuracies range approximately from \(10^{-4}\) to \(10^{-3}\) per site.
We therefore write
\begin{equation}
    \epsilon_E(L)=f_L L^2,
    \label{eq:lattice_energy_tolerance}
\end{equation}
where \(f_L\) is a lattice-dependent per-site energy-error factor. The values
of \(f_L\) are given in \cref{tab:circuit_parameters}. The corresponding total
energy tolerance is multiplied by \(L^2\) because an \(L\times L\) lattice
contains \(L^2\) sites. The phase-register size is recalculated for each
lattice size rather than being fixed across all simulations.

\subsubsection{Rotation-synthesis error}

\noindent
The arbitrary rotations appearing in the circuit must be approximated by
fault-tolerant gate sequences. Let \(\epsilon_{\mathrm{RS}}(L)\) denote the
synthesis tolerance assigned to each rotation for lattice size \(L\), and let
\(N_{\mathrm{rot}}(L)\) be the total number of synthesized rotations. Using a
conservative additive bound, the total rotation-synthesis error satisfies
\begin{equation}
    \epsilon_{\mathrm{RS,tot}}(L)
    \leq
    \sum_{j=1}^{N_{\mathrm{rot}}(L)}
    \epsilon_{\mathrm{RS},j}
    =
    N_{\mathrm{rot}}(L)\epsilon_{\mathrm{RS}}(L),
    \label{eq:total_rotation_synthesis_error}
\end{equation}
when the same tolerance is used for every rotation.

Because the total synthesis error depends on the number of rotations in the
circuit, a single per-rotation tolerance does not provide the same total error
bound for every lattice size. We therefore select
\(\epsilon_{\mathrm{RS}}(L)\) separately for each lattice size. The
phase-register and rotation-synthesis settings are chosen jointly so that
\begin{equation}
    \epsilon_{\mathrm{phase}}(L)
    +
    \epsilon_{\mathrm{RS,tot}}(L)
    \leq 0.05
    \label{eq:combined_algorithmic_error_constraint}
\end{equation}
for every simulated circuit.

\begin{table}[t]
    \centering
    \small
    \begin{tabular}{c c c c}
        \toprule
        \(L\)
        & \(f_L=\epsilon_E/L^2\)
        & \(\epsilon_E=f_LL^2\)
        & \(\epsilon_{\mathrm{RS}}\) \\
        \midrule
        2  & \(1\times10^{-3}\) & \(4.0\times10^{-3}\)  & \(1\times10^{-4}\) \\
        4  & \(1\times10^{-3}\) & \(1.6\times10^{-2}\)  & \(1\times10^{-4}\) \\
        6  & \(1\times10^{-3}\) & \(3.6\times10^{-2}\)  & \(2\times10^{-8}\) \\
        8  & \(7\times10^{-4}\) & \(4.48\times10^{-2}\) & \(1\times10^{-5}\) \\
        10 & \(4\times10^{-4}\) & \(4.0\times10^{-2}\)  & \(9\times10^{-9}\) \\
        20 & \(1\times10^{-4}\) & \(4.0\times10^{-2}\)  & \(2\times10^{-9}\) \\
        \bottomrule
    \end{tabular}
    \caption{
        Algorithmic-error parameters used for each Fermi--Hubbard lattice
        size. The per-site factor \(f_L\) sets the total energy tolerance
        \(\epsilon_E=f_LL^2\), which is used to determine the number of phase
        qubits through \cref{eq:number_of_phase_qubits}. The quantity
        \(\epsilon_{\mathrm{RS}}\) is the approximation tolerance assigned to
        each synthesized rotation. Energy tolerances are given in units where
        \(t=1\). The phase-register and rotation-synthesis settings are
        selected jointly so that their estimated total contribution to the
        algorithmic error is at most \(5\%\).
    }
    \label{tab:circuit_parameters}
\end{table}

\subsection{Simplifying subcircuits to common gates}

\noindent This circuit is made up of 2 subcircuits: the controlled unitary and the inverse quantum Fourier transform (QFT$^\dagger$). In this subsection we will elucidate how the subcircuits were compiled down to simple gates such as multi-qubit Toffolis  and rotations. See~\cref{tab:circuit_parameters} for a summary of the parameters used to create each of the subcircuits.

\subsubsection{Controlled Unitary}

\noindent The controlled unitary contains all the information for the Hamiltonian. One can write the Fermi-Hubbard Hamiltonian as
\begin{align}
    H = tH_K + UH_P
\end{align}
a kinetic part $H_K$ and a potential part $H_P$ each with respective coefficients $t$ and $U$. We set $t=1$ and $U=8$, a commonly studied parameter regime in the literature~\cite{babbush2018encoding}. We used the Jordan-Wigner fermion-to-qubit mapping to map the Hamiltonian to a sum of Pauli gates which could then be implemented as an LCU using \texttt{SELECT} and \texttt{PREP} gates following the prescription of~\cite{babbush2018encoding} and using the circuit improvements shown in~\cite{kan2025resource}. A description of how the \texttt{SELECT} and \texttt{PREP} are implemented is given in~\cref{fig:full_test_circuit}.

We expect the active volume of this portion of the circuit to be dominated by Toffoli gates. It's clear to see from~\cref{fig:full_test_circuit} that Toffolis are very common gates in each of the smaller subcircuits of the controlled unitary. Any remaining multi-controlled gates that aren't explicitly decomposed in~\cref{fig:full_test_circuit} were turned into single controlled gates, using left and right elbows~\cite{gidney2018halving}.

\subsubsection{Inverse QFT}

We chose to decompose the quantum Fourier transform using the standard method of Hadamards interleaved with phase gradient gates. These phase gradient gates decompose down into a series of controlled phase gates of smaller and smaller angles. The only special angle that we optimized for was the controlled S gate. All others were treated as arbitrary angles. Thus, one would expect that the cost of the QFT$^\dagger$ is dominated by rotations.

\subsection{Validity of scheduling by parts}
\label{sec:validity_of_parts}

\noindent For large circuits, constructing and scheduling a single monolithic DAG can be computationally expensive and memory intensive. We therefore often use a component-wise approximation, which we refer to as scheduling by parts. In this approach, the circuit is decomposed into repeated sub-circuit components, each component is converted into a DAG and scheduled independently, and the resulting schedule length is multiplied by the number of times that component appears in the full circuit. If component $i$ has schedule length $L_i$ and multiplicity $m_i$, the total schedule length is approximated as~\cref{eq:parts_schedule_length}.
There are two problems with parting out a circuit: firstly, stale state dependencies are lost between components, underestimating their impact on reaction limited behavior. Secondly, the final logical cycle of each components schedule can be artificially empty, gates from the next component could occupy this otherwise unused space, whereas the component-wise approximation repeats the partially empty final cycle each time the component is used. This effect tends to overestimate the runtime, and becomes more important when each component contains fewer logical cycles. We therefore use the ratio of final logical cycles over total logical cycles, \cref{eq:boundary_threshold}, to determine the validity of our by parts runtime. But how do we decide a good threshold ratio?

In this section we investigate the validity of scheduling a quantum circuit by parts rather than as a single monolithic DAG. We use a $4\times4$ lattice test circuit with halved phase precision—reduced so that constructing the full QPE DAG remains tractable on a laptop—and schedule it both by parts and in full across a range of computer sizes. We then identify the computer sizes at which the runtime difference between the two methods exceeds 5\%. The $\epsilon_{\mathrm{bdry}}$ of the last data point that remains within this 5\% cutoff defines our upper bound on valid $\epsilon_{\mathrm{bdry}}$  for the parts approximation.

\begin{figure}
    \centering
    \includegraphics[width=\linewidth]{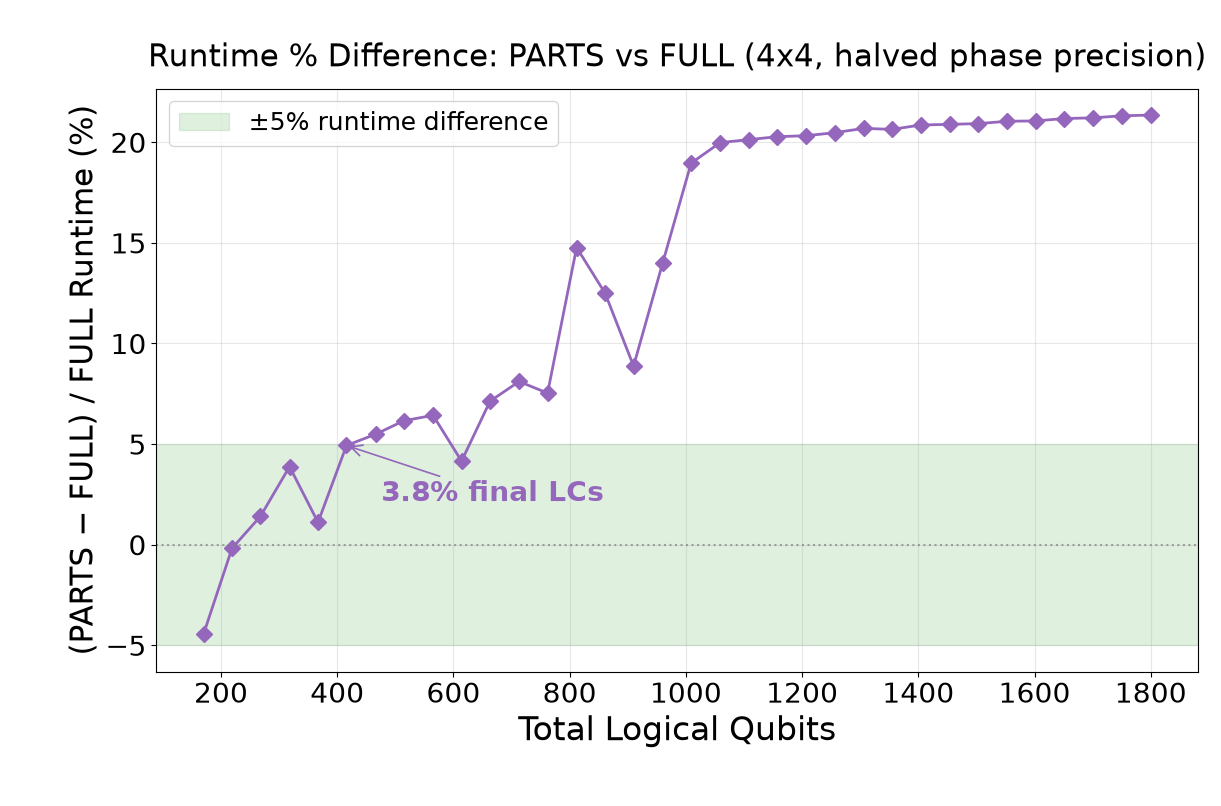}
    \caption{
    Signed percentage difference between the runtime obtained by scheduling a reduced-precision $4\times4$ Fermi--Hubbard circuit by parts or as a single full DAG. Positive values indicate that the parts approximation overestimates the full-circuit runtime. The approximation remains within $5\%$ up to approximately 420 logical qubits, corresponding to $\epsilon_{\mathrm{bdry}}=3.8\%$, which we use as the validity threshold for the parts scheduling method.
    }
    \label{fig:parts_vs_full_pct_diff}
\end{figure}

In~\cref{fig:parts_vs_full_pct_diff}, we see how the percentage difference in runtime between the combined parts and full schedule changes with computer size. There is an upward trend with qubit count, which is expected as the large computers have large workspaces, this means each component has fewer logical cycles and the impact of the final logical cycle of component schedules matter more. The 5\% difference holds until around 420 qubits, where $\epsilon_{\mathrm{bdry}}=3.80\%$. So we use $\epsilon_{\mathrm{bdry}}=3.80\%$ as a working threshold for when parts scheduling gives runtimes within $5\%$ of the corresponding full circuit schedule. At small computer sizes, below approximately 200 logical qubits, the component-wise schedule predicts a slightly faster runtime than the full-circuit schedule. Since boundary effects would normally make the parts schedule longer, this behavior suggests that the greedy scheduler can sometimes find a more efficient ordering when applied to smaller component DAGs than when applied to the full DAG. We do not rely on this effect, and leave a more detailed investigation of the interaction between DAG size and greedy scheduling performance to future work.

\begin{figure}[!tbp]
    \centering
    \includegraphics[width=\linewidth]{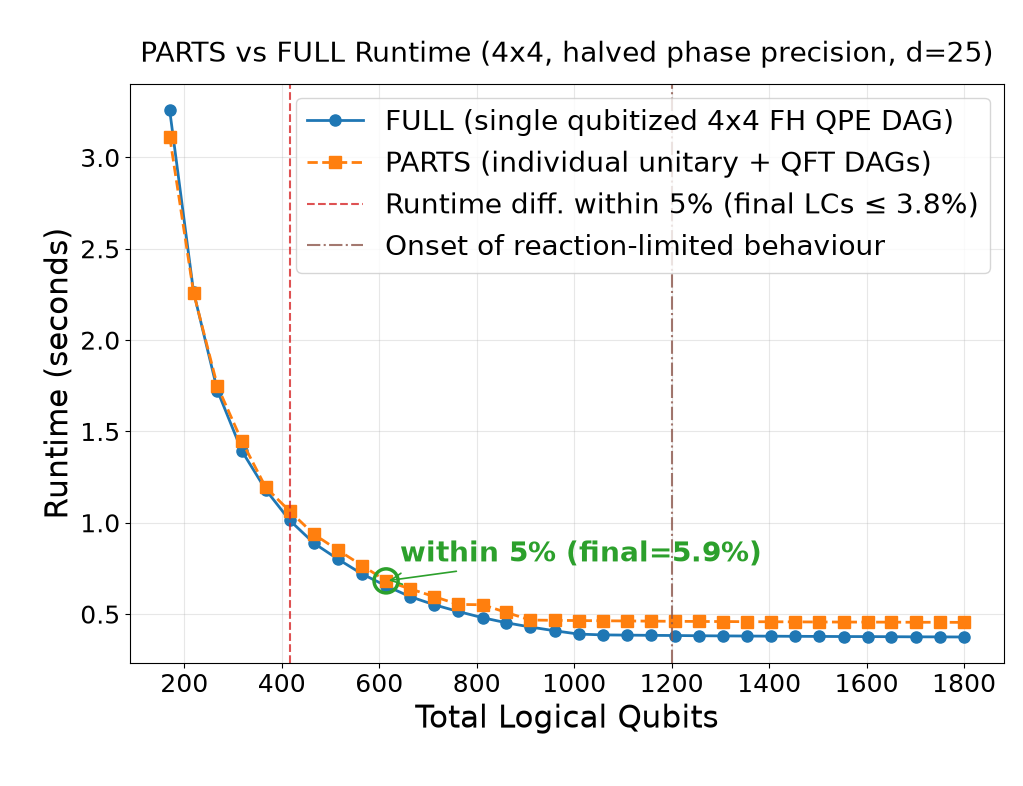}
    \caption{
    Runtime comparison between parts scheduling and full-DAG scheduling for the reduced-precision $4\times4$ Fermi--Hubbard circuit. The two methods agree closely below the threshold identified in~\cref{fig:parts_vs_full_pct_diff}. At larger computer sizes, the parts approximation generally overestimates the runtime because each repeated component contributes a final logical cycle that may be less efficiently packed than in the full-circuit schedule. At point around 600 qubits, where the final LCs make up 5.9\% of all LCs, falls back into the 5\% runtime difference range. This point happens to have a well filled final cycle and is still in the non-reaction limited regime where stale stales lost at the boundary have a minimal effect on runtime.
    }
    \label{fig:parts_vs_full_runtime}
\end{figure}

In~\cref{fig:parts_vs_full_runtime}, we show the corresponding runtimes for both scheduling methods. Below the validity threshold, the two methods agree closely. Above this threshold, the component-wise approximation generally remains reasonable, but tends to overestimate the runtime due to inefficient packing of each components final cycle. Thus, scheduling by parts can still be useful beyond the strict threshold when a modest runtime overestimate is acceptable. There is an isolated point near 600 logical qubits where the component-wise runtime again lies within $5\%$ of the full circuit runtime, despite being beyond the threshold identified above. This occurs because the final logical cycle of the relevant component schedule happens to be relatively well filled, reducing the boundary-cycle error. We therefore interpret $\epsilon_{\mathrm{bdry}}=3.9\%$ as a conservative threshold for reliable component-wise scheduling rather than as a sharp boundary. Beyond the 1200 qubit mark we see a greater disagreement in runtime, this is because we now have reaction limited behavior which this lost at the boundary of component schedules.

\subsection{Simplifying single controls}

\begin{figure}[b]
    \centering
    \includegraphics[width=.7\columnwidth]{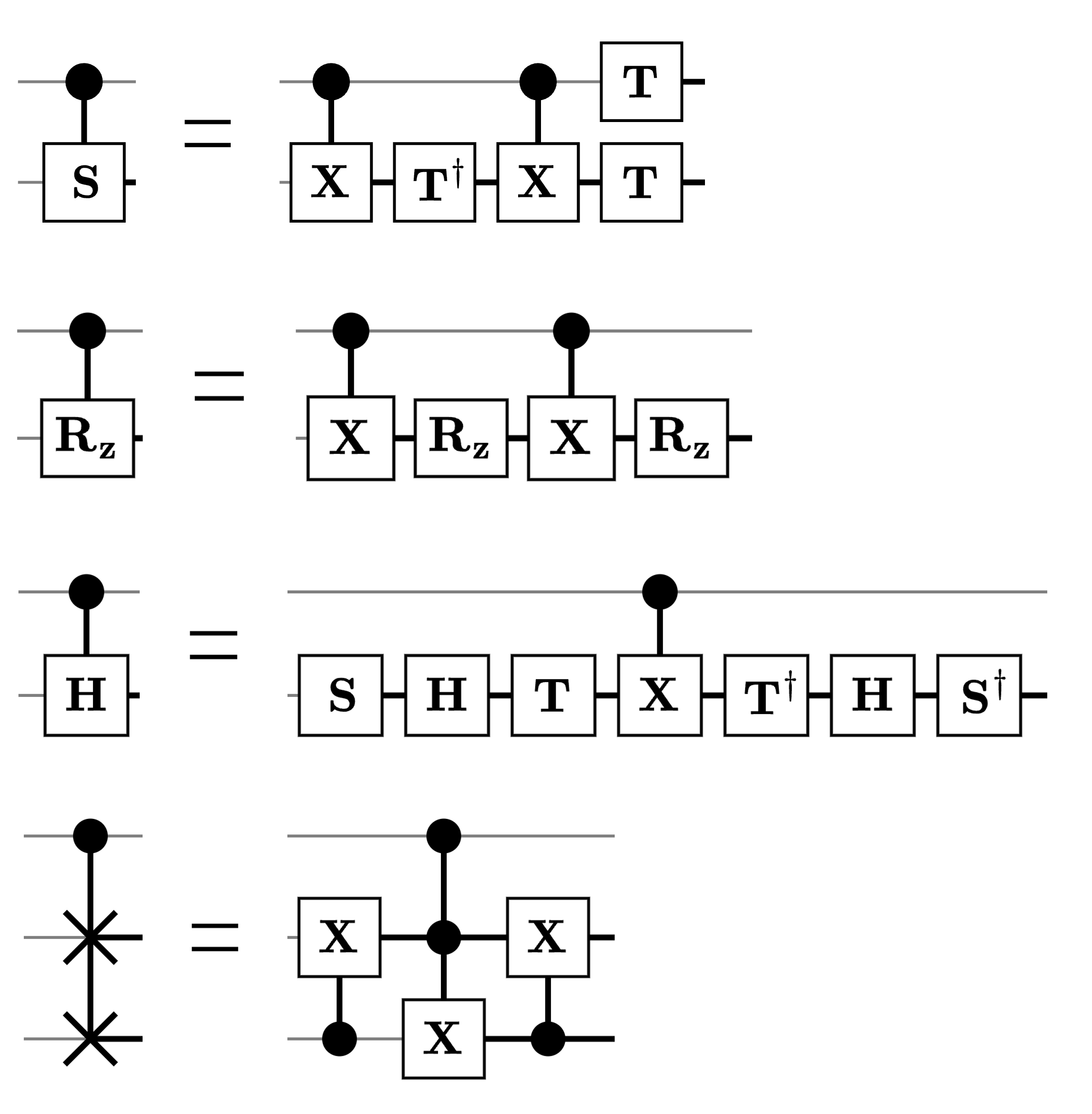}
    \caption{A summary of the single control decompositions that were employed to break down all the gates in~\cref{fig:full_test_circuit}. As in previous figures, we are only concerned with the costs of the gates, so we omit the particular angle which an arbitrary RZ is done. The controlled S and controlled RZ decompositions were employed in the phase gradient circuit. In that case controlled T gates were treated as if they were arbitrary angle rotations. The controlled Hadamard and controlled swaps are used in the controlled unitary.}
    \label{fig:single_control_decompositions}
\end{figure}

\noindent Although~\cref{fig:full_test_circuit} goes a long way towards breaking down large gates into more familiar ones, we still need to further decompose our gates into ones where the AV and number of reaction layers are known. To this end we employed several gate decompositions illustrated in~\cref{fig:single_control_decompositions}. To implement arbitrary rotations, we employed the Ross-Selinger~\cite{RossSelinger2016Optimal} technique to decompose them into sequences of Hadamards, phase gates, and T gates. These decompositions were enough to reduce all gates in the circuit to gates whose AVs were listed in Table 1 of~\cite{Litinski22Active}.

\end{document}